\documentclass[]{ant_tech_report}

\usepackage{latexsym}
\usepackage[utf8]{inputenc}
\usepackage{inconsolata}
\ifdefined\pdfimageresolution
\fi
\usepackage{siunitx}
\usepackage{url}
\usepackage{multicol}
\usepackage{colortbl}
\usepackage{wrapfig}
\usepackage{makecell}
\usepackage{adjustbox}
\usepackage{algorithm}
\usepackage[noend]{algpseudocode}
\usepackage{fancyvrb}
\usepackage{fvextra}
\usepackage{soul}
\usepackage{float}
\usepackage{tikz}
\usepackage{fontawesome5}
\titleformat*{\paragraph}{\bfseries\itshape}

\definecolor{memguiblue}{HTML}{E8F1FA}
\definecolor{cvprblue}{rgb}{0.21,0.49,0.74}
\definecolor{memorycolor}{RGB}{216, 228, 244}
\definecolor{foldcolor}{RGB}{229, 190, 217}
\definecolor{stepcolor}{RGB}{232, 204, 204}
\definecolor{memguigreen}{RGB}{71, 172, 20}
\definecolor{memguired}{RGB}{238, 68, 51}
\definecolor{toolblue}{HTML}{263D4A}
\definecolor{histcolor}{HTML}{B79AD0}
\definecolor{recentcolor}{HTML}{E46F6F}
\definecolor{memstatecolor}{HTML}{F2A65A}
\definecolor{codegray}{gray}{0.95}

\newenvironment{paperresources}
  {\par\noindent\begin{minipage}{0.92\textwidth}
    \centering\footnotesize\sffamily\color{black!82}
    \setlength{\parindent}{0pt}\setlength{\parskip}{2pt}}
  {\end{minipage}\par\vspace{1.7mm}}
\newcommand{\resourceprojecticon}{\textcolor{antblue}{\faGlobe}}
\newcommand{\resourcegithubicon}{\textcolor{antblue}{\faGithub}}
\newcommand{\resourcehficon}{%
  \raisebox{0.3ex}{%
    \tikz[baseline=-0.5ex,scale=0.105]{%
      \fill[antblue] (0,0) circle (1);
      \fill[antblue] (-0.8,0.68) circle (0.34);
      \fill[antblue] (0.8,0.68) circle (0.34);
      \fill[white] (-0.34,0.22) circle (0.11);
      \fill[white] (0.34,0.22) circle (0.11);
      \draw[white,line width=0.12pt,line cap=round] (-0.36,-0.26) .. controls (0,-0.55) .. (0.36,-0.26);
    }%
  }%
}
\newcommand{\paperresourceicon}[4]{%
  #1\ \textbf{#2:}\ \href{#3}{{\ttfamily\fontsize{8}{9}\selectfont #4}}%
}

\newcommand{\obsbox}[1]{%
  \begin{tcolorbox}[colframe=black!50, colback=cvprblue!8, boxrule=1.2pt, arc=2mm,
    top=4pt, bottom=4pt, left=6pt, right=6pt, boxsep=2pt, fontupper=\itshape]
    #1
  \end{tcolorbox}
}

\tcbset{
  promptstyle/.style={
    colback=codegray,
    colframe=black,
    fontupper=\ttfamily\scriptsize,
    fonttitle=\ttfamily\scriptsize\raggedright,
    coltitle=white,
    colbacktitle=black,
    sharp corners,
    boxrule=0.5pt,
    enhanced,
    breakable,
    left=4pt,
    right=4pt,
    top=4pt,
    bottom=4pt,
  }
}

\newcounter{insight}
\newcommand{\insight}[1]{%
  \refstepcounter{insight}%
  \vspace{0.30em}
  \noindent
  \begingroup
  \setlength{\fboxsep}{0pt}%
  \colorbox{histcolor!7}{%
    \parbox{0.97\columnwidth}{%
      \vspace{0.35em}
      \hspace{0.65em}%
      \textcolor{histcolor}{\rule{1.4pt}{1.05em}}%
      \hspace{0.55em}%
      \parbox[t]{0.89\columnwidth}{%
        \textbf{\textsc{Insight}~\theinsight.}~\emph{#1}%
      }%
      \vspace{0.35em}
    }%
  }%
  \endgroup
  \vspace{0.30em}
}

\newcommand{\memaction}[1]{\textcolor{toolblue}{\texttt{#1}}}
\newcommand{\foldop}[1]{\textcolor[HTML]{9B6BB3}{\texttt{#1}}}

\newcommand{\toolfield}[1]{\textcolor{memstatecolor}{\texttt{#1}}}
\newcommand{\recentfield}[1]{\textcolor{recentcolor}{\texttt{#1}}}

\newcommand{\symmem}{$\spadesuit$\xspace}
\newcommand{\symfold}{$\clubsuit$\xspace}
\newcommand{\symstep}{$\diamondsuit$\xspace}

\definecolor{hlbestcolor}{HTML}{CFE2FF}
\definecolor{hlsecondcolor}{HTML}{E8F1FF}
\definecolor{oursrowcolor}{HTML}{F4EEF8}
\definecolor{oursrowcolorsecond}{HTML}{F9F2FB}

\newcommand{\llmname}[1]{{\fontfamily{pcr}\selectfont{#1}}}

\newcommand{\ourmethod}{MemGUI-Agent\xspace}
\newcommand{\conact}{\textsc{ConAct}\xspace}
\definecolor{risecolor}{HTML}{159A9C}
\newcommand{\rise}[1]{\textcolor{risecolor}{#1}}
\newcommand{\drop}[1]{\textcolor{red!75!black}{#1}}

\newcommand{\includefig}[2][width=\linewidth]{\IfFileExists{#2}{\includegraphics[#1]{#2}}{\centering\fbox{\parbox{\dimexpr\linewidth-2\fboxsep-2\fboxrule}{\centering\small [Placeholder: \texttt{#2}]}}}}

\title{MemGUI-Agent: An End-to-End Long-Horizon Mobile GUI Agent with Proactive Context Management}

\author{%
\parbox{\textwidth}{\centering
Guangyi Liu$^{1,2}$,
Gao Wu$^{1,2}$,
Congxiao Liu$^{1,2}$,
Pengxiang Zhao$^{1,2}$,
Liang Liu$^{1}$,
Mading Li$^{2,\dagger}$\\[1mm]
Zhang Qi$^{2}$,
Mengyan Wang$^{2}$,
Liang Guo$^{2}$,
Yong Liu$^{1,\S}$
}}

\affiliation{%
\parbox{\textwidth}{\centering\small
$^1$Zhejiang University \quad $^2$Kuaishou Technology
}}

\renewcommand{\contributionlist}{\vspace{-2mm}}
\newcommand{\authorfootnotes}{%
{\scriptsize
\begin{tabular}{@{}l@{}}
\rule{34mm}{0.4pt}\\[-0.2ex]
$^\dagger$Project lead.\quad $^\S$Corresponding author.
\end{tabular}}}
\fancypagestyle{firststyle}{
    \fancyhf{}
    \fancyhead[R]{
        \vskip 3mm
    \includegraphics[height=13mm]{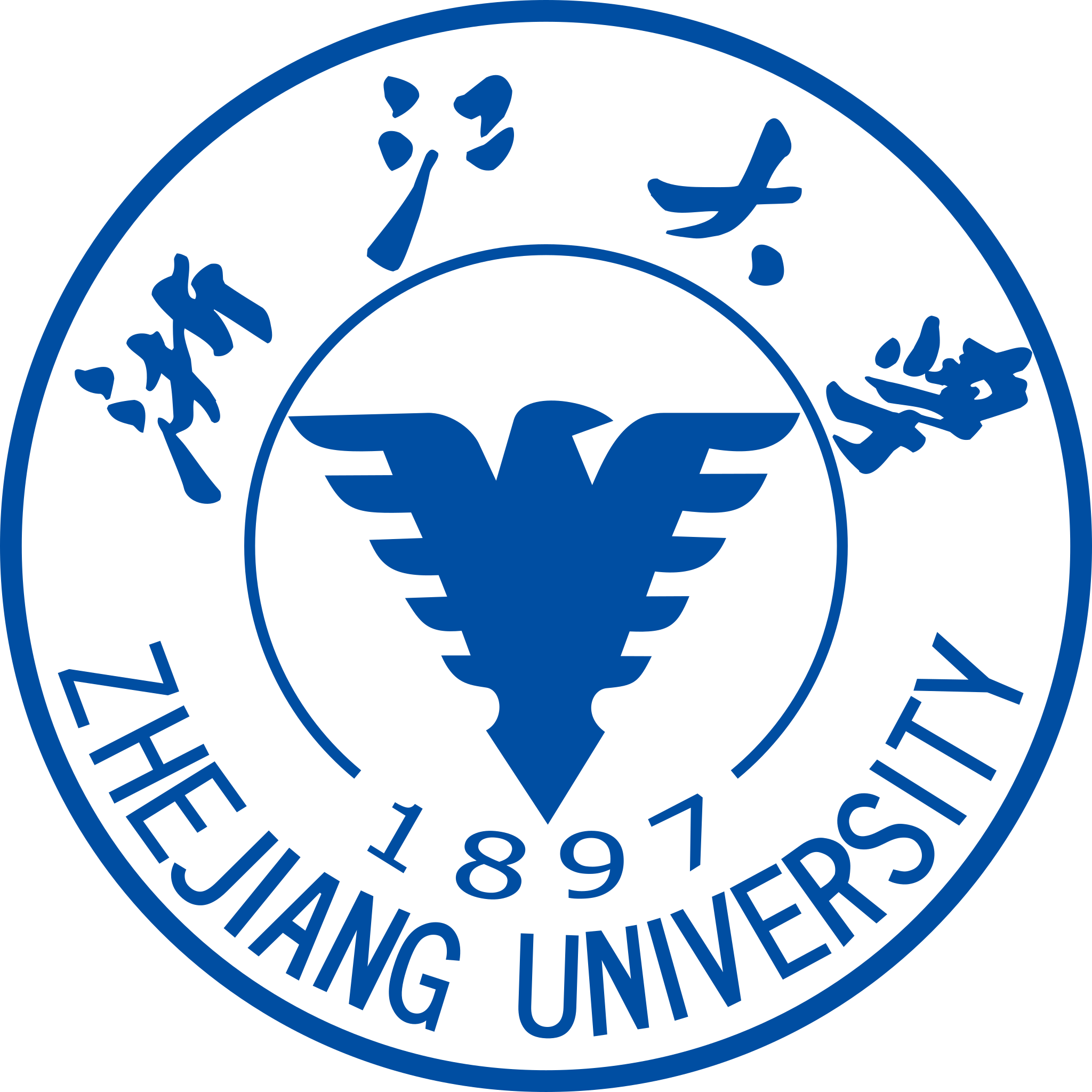}
    }
    \fancyhead[L]{
        \vskip 3mm
    \includegraphics[height=10mm]{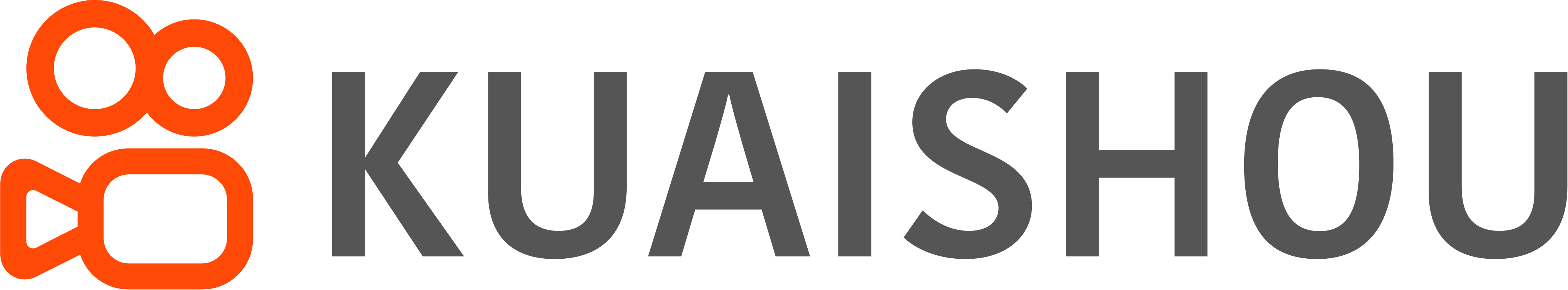}
    }
    \fancyfoot[L]{\raisebox{1.15\baselineskip}[0pt][0pt]{\authorfootnotes}}
    \fancyfoot[C]{\thepage}
}
\renewcommand{\checkdatalist}{%
\begin{paperresources}
\paperresourceicon{\resourceprojecticon}{Project Page}{https://memgui-agent.github.io/}{memgui-agent.github.io}
\quad\quad
\paperresourceicon{\resourcehficon}{Dataset}{https://huggingface.co/datasets/lgy0404/MemGUI-3K}{lgy0404/MemGUI-3K}\\
\paperresourceicon{\resourcegithubicon}{Code}{https://github.com/kwai/MemGUI-Agent}{github.com/kwai/MemGUI-Agent}
\quad\quad
\paperresourceicon{\resourcehficon}{Model}{https://huggingface.co/lgy0404/MemGUI-8B-SFT}{lgy0404/MemGUI-8B-SFT}
\end{paperresources}}

\abstract{MLLM-based mobile GUI agents have made substantial progress on short-horizon tasks, yet remain unreliable on long-horizon tasks that require retaining intermediate facts across many steps and app transitions. We attribute this limitation to ReAct-style prompting, which passively accumulates per-step records, leading to prompt explosion and dilution of critical cross-app facts.
To address this, \textbf{\emph{(i)}} we introduce \textbf{\ourmethod}, an end-to-end long-horizon mobile GUI agent with proactive context management. \ourmethod is built on \textbf{\underline{Con}}text-as-\textbf{\underline{Act}}ion (\textbf{\conact}), which casts context management as first-class actions emitted by the same policy that selects UI actions. Instead of passively appending history, \conact maintains three structured context fields: folded action history, folded UI state, and recent step record, preserving critical UI facts while keeping context compact.
To make proactive context management learnable across model scales, \textbf{\emph{(ii)}} we construct \textbf{MemGUI-3K}, a 2{,}956-trajectory dataset with full \conact annotations for supervised training and offline analysis. \textbf{\emph{(iii)}} Training an 8B model on MemGUI-3K produces \textbf{MemGUI-8B-SFT}, an 8B \ourmethod that achieves the best open-data 8B performance on MemGUI-Bench and generalizes to the out-of-distribution MobileWorld benchmark. Code, data, and trained models will be released at \url{https://memgui-agent.github.io/}.
}

\begin{document}
\maketitle
\enlargethispage{6\baselineskip}
\begingroup
\vspace*{-2.0em}
\captionsetup{font=footnotesize}
\noindent\makebox[\textwidth][c]{%
\begin{minipage}{0.99\textwidth}
  \centering
  \includegraphics[width=0.99\linewidth]{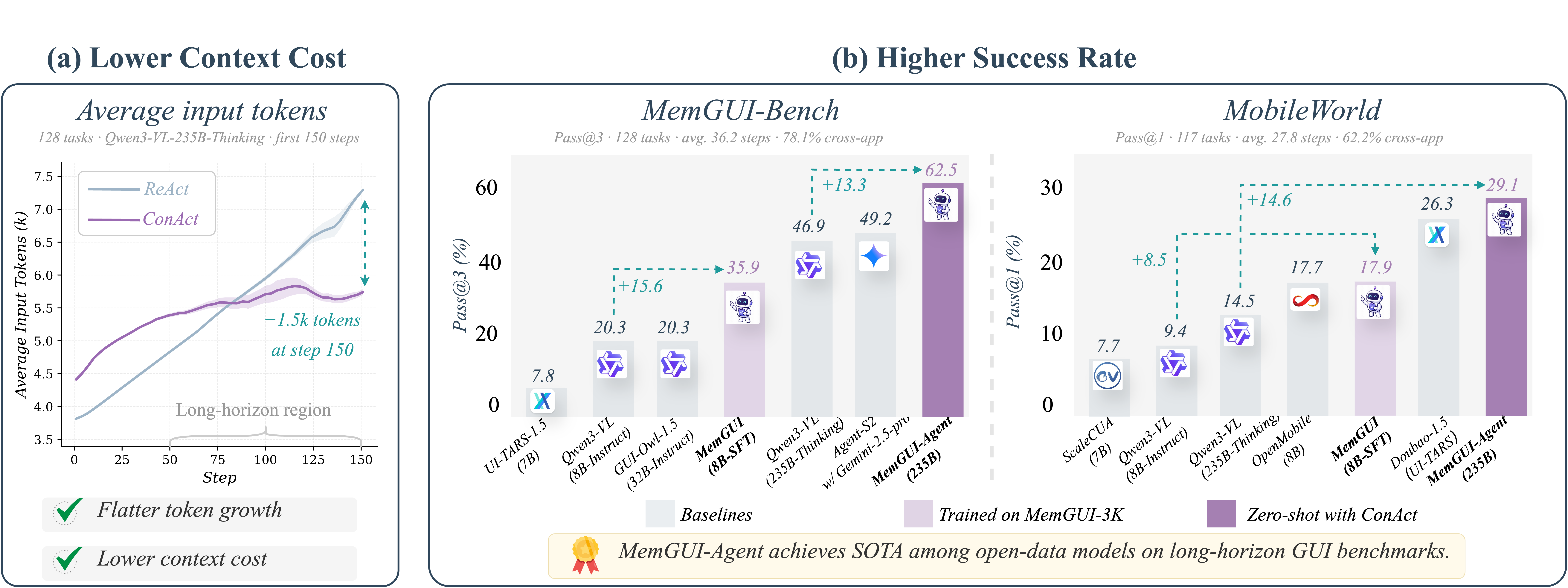}
  \vspace{-0.5em}
  \captionof{figure}{
  \textbf{Context efficiency and benchmark performance of \ourmethod.}
  \textbf{(a)}~Pass@3 token trajectories on 128 MemGUI-Bench tasks, truncated to 150 steps. Purple/blue lines denote \conact/ReAct; translucent bands show per-step uncertainty across trajectories. \conact saves $\sim$1.5k input tokens by step 150.
  \textbf{(b)}~Benchmark results on MemGUI-Bench and MobileWorld, where \ourmethod improves both 235B zero-shot and 8B-SFT settings, with MemGUI-8B-SFT setting the best open-data 8B performance.
  }
  \label{fig:main-performance}
\end{minipage}}
\vspace{-0.8em}
\endgroup

\clearpage

\section{Introduction}
\label{sec:intro}
\vspace{-0.3em}

Recent MLLM-based mobile GUI agents can understand screenshots, reason over user goals, and control devices~\cite{gao2026ui,zhou2025mai,hong2024cogagent,xu2026mobile,liu2025llm}. However, they remain unreliable in long-horizon settings that require retaining intermediate facts, tracking multi-step progress, and carrying information across app boundaries~\cite{liu2026memgui,kong2025mobileworld}. As task horizons grow, even the strongest 8B end-to-end model, \llmname{GUI-Owl-1.5-8B}~\cite{xu2026mobile}, drops from 71.6\% on AndroidWorld~\cite{rawles2025androidworld} (avg.\ 8.4 steps) to 38.2\% on MobileWorld~\cite{kong2025mobileworld} (avg.\ 27.8 steps), and further to 11.7\% on MemGUI-Bench~\cite{liu2026memgui} (avg.\ 36.2 steps).

\begin{wrapfigure}{r}{0.52\textwidth}
  \vspace{-1.0em}
  \centering
  \captionsetup{font=footnotesize}
  \includegraphics[width=\linewidth]{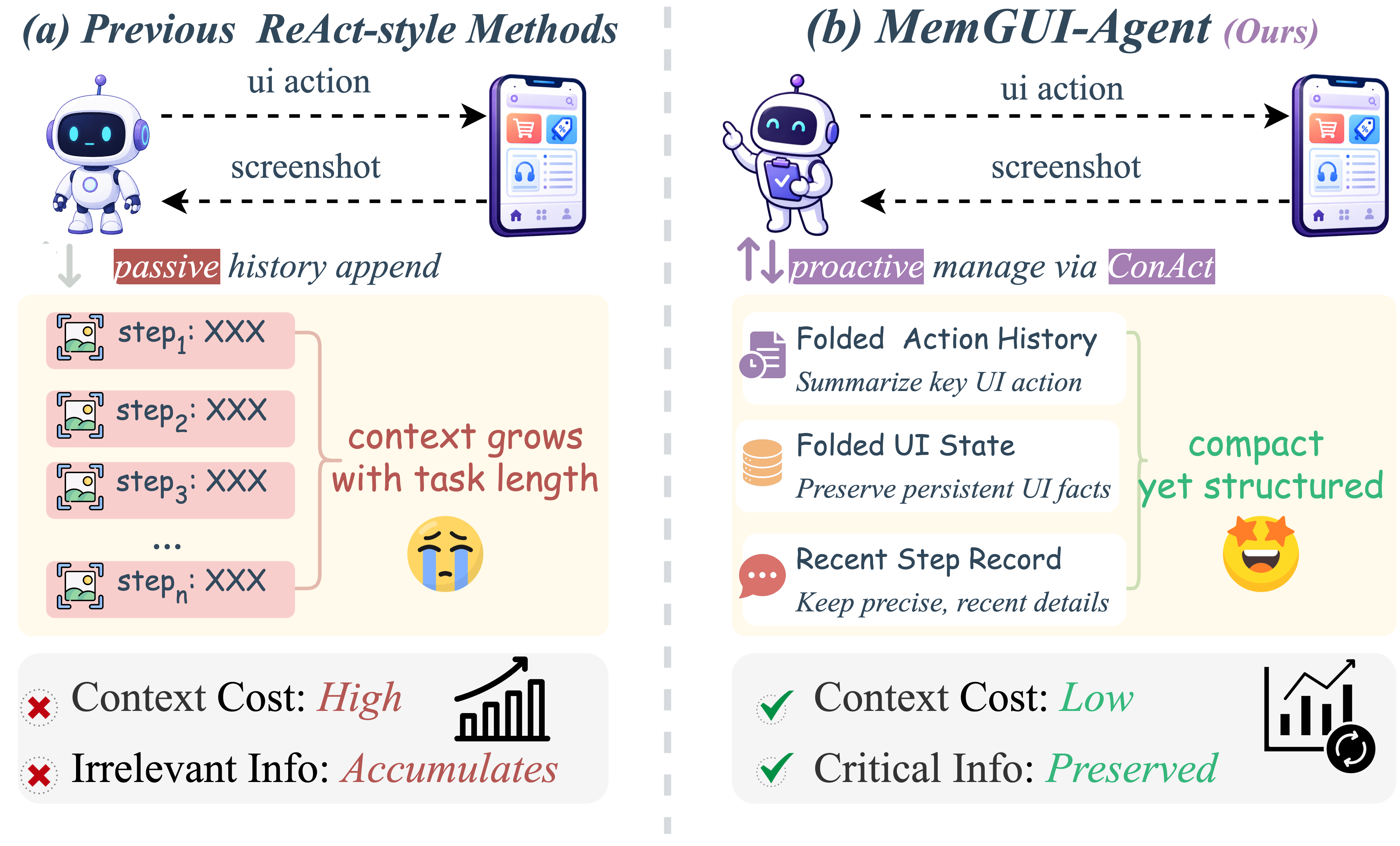}
  \vspace{-0.6em}
  \caption[Context management in mobile GUI agents.]{
    Context management in mobile GUI agents.
    \textbf{(a)}~Previous ReAct-style methods passively append per-step records, making context grow with task length and accumulate noise.
    \textbf{(b)}~\ourmethod proactively manages context via \conact's three structured fields:
    \textcolor{histcolor}{Folded Action History},
    \textcolor{memstatecolor}{Folded UI State},
    and \textcolor{recentcolor}{Recent Step Record},
    keeping context compact while preserving critical UI facts.
    }
  \label{fig:teaser}
  \vspace{-1.2em}
\end{wrapfigure}

We attribute this degradation to context management. \emph{Agentic frameworks} couple a backbone MLLM with planning, memory, and grounding modules~\cite{agashe2025agent,rawles2025androidworld,wang2024mobile,wang2025mobileagentselfevolving}, but rely on complex pipelines and often proprietary backbones. \emph{End-to-end models} are simpler and more deployable~\cite{gao2026ui,zhou2025mai,xu2026mobile}, yet manage context through passive strategies (e.g., Action-Thought, Multi-turn, Rule-based, No History; see Figure~\ref{fig:teaser} and Table~\ref{tab:main-results}) that append, truncate, or discard records mechanically rather than according to task needs. Consequently, prompts grow with task horizon (\emph{prompt explosion}), while critical cross-app facts such as prices, identifiers, or copied text may be diluted, paraphrased, truncated, or forgotten (\emph{information loss})~\cite{liu2026memgui}. We review related mobile GUI agents and context-management methods in Appendix~\ref{sec:related-work}.

Long-horizon mobile GUI control requires both compact working context and persistent UI-derived facts across screens, steps, and app transitions. Existing strategies fail to provide both: external memory moves context curation outside the end-to-end policy, while prompt-based or rule-based methods either accumulate passive logs or discard old facts mechanically. This motivates a policy-level mechanism that decides what to compress, what to remember, and what to keep available while acting.

\obsbox{\textit{Can a single end-to-end mobile GUI agent jointly choose UI actions and manage its working context, so that long-horizon interaction stays compact while critical UI facts persist across steps and app transitions?}}

We answer this question with \textbf{\ourmethod}, an end-to-end long-horizon mobile GUI agent with proactive context management. \ourmethod is built on \textbf{\underline{Con}}text-as-\textbf{\underline{Act}}ion (\textbf{\conact}), which casts context management as first-class actions emitted alongside UI actions. Instead of passively appending history, \conact maintains three structured context fields: \textcolor{histcolor}{\emph{Folded Action History}} \textcolor{histcolor}{$H_t$} for compressed trajectory summaries, \textcolor{memstatecolor}{\emph{Folded UI State}} \textcolor{memstatecolor}{$M_t$} for persistent UI-derived facts, and \textcolor{recentcolor}{\emph{Recent Step Record}} \textcolor{recentcolor}{$L_t$} for recent interaction details. All three are predicted in a single forward pass, keeping context curation inside the agent policy.

As shown in Figure~\ref{fig:main-performance}(a), \conact keeps prompt growth flatter than ReAct-style prompting. Figure~\ref{fig:main-performance}(b) shows that this lower context cost translates into stronger long-horizon task success: applied zero-shot to \llmname{Qwen3-VL-235B-Thinking}, \ourmethod reaches 62.5\% Pass@3 on MemGUI-Bench, surpassing agentic workflows built on \llmname{Gemini-2.5-Pro}. To make proactive context management learnable across model scales, we construct \textbf{MemGUI-3K}, a 2{,}956-trajectory dataset with full \conact annotations. Training an 8B model on MemGUI-3K produces \textbf{MemGUI-8B-SFT}, which achieves the best open-data 8B performance on MemGUI-Bench and generalizes to the out-of-distribution MobileWorld benchmark.

\paragraph{Contributions.}
\textbf{(1)}~We introduce \textbf{\ourmethod}, an end-to-end long-horizon mobile GUI agent built on \conact, a Context-as-Action paradigm that unifies history folding, UI memory, and self-describing step outputs within a single policy~(\S\ref{sec:method}).
\textbf{(2)}~We construct \textbf{MemGUI-3K}, a 2{,}956-trajectory dataset with full \conact annotations, supporting supervised training and offline analysis of proactive context management across model scales~(\S\ref{sec:dataset}).
\textbf{(3)}~We train \textbf{MemGUI-8B-SFT}, an 8B \ourmethod supervised on MemGUI-3K, achieving the best open-data 8B performance on MemGUI-Bench and generalizing to the out-of-distribution MobileWorld benchmark~(\S\ref{sec:experiments}).

\begin{figure*}[t]
  \centering
  \includefig[width=0.92\textwidth]{images/contexact/contextact-framework.drawio.png}
  \caption{Single-step execution of \ourmethod with \conact. The MLLM $\pi_\theta$ takes screenshot $I_t$ and structured state $\mathcal{S}_t = (G, \textcolor{histcolor}{H_t}, \textcolor{memstatecolor}{M_t}, \textcolor{recentcolor}{L_t})$ and emits five outputs $(\tau_t, \phi_t, a_t, o_t, \iota_t)$. Three parallel updates follow: \textbf{(A)}~folding directive $\phi_t$ updates \textcolor{histcolor}{$H_{t+1}$}; \textbf{(B)}~tool call $a_t$ acts on the environment or updates \textcolor{memstatecolor}{$M_{t+1}$}; \textbf{(C)}~recent step record \textcolor{recentcolor}{$L_{t+1}$} is assembled from $(o_t, \iota_t, a_t, r_t)$. The next state $\mathcal{S}_{t+1}$ feeds into step $t{+}1$.}
  \label{fig:framework}
\end{figure*}

\section{\ourmethod: End-to-End Mobile GUI Agent with \conact}
\label{sec:method}
\vspace{-0.25em}

\conact makes context management a first-class part of the action policy in an end-to-end mobile GUI agent. Instead of treating context as a passive log outside the model's control, the agent jointly decides \emph{what UI or memory action to execute}, \emph{what history to fold}, and \emph{how to describe the current interaction}. This turns context maintenance into policy-level behavior. Figure~\ref{fig:framework} illustrates the single-step execution flow.

\paragraph{Problem formulation.}
We formulate mobile GUI automation as a sequential decision problem with a structured working context. Given task goal $G$ and screenshot $I_t$, the agent observes
\begin{equation}
\label{eq:state}
\mathcal{S}_t =
(G,\textcolor{histcolor}{H_t},
\textcolor{memstatecolor}{M_t},
\textcolor{recentcolor}{L_t}),
\end{equation}
where \textcolor{histcolor}{$H_t$}, \textcolor{memstatecolor}{$M_t$}, and \textcolor{recentcolor}{$L_t$} denote \emph{Folded Action History}, \emph{Folded UI State}, and \emph{Recent Step Record}, storing compressed trajectory summaries, persistent UI-derived facts, and latest-step details. At each step, the MLLM policy emits a joint \conact decision:
\begin{equation}
\label{eq:joint-output}
\begin{aligned}
y_t
&= (\tau_t,\textcolor{histcolor}{\phi_t},
\textcolor{memstatecolor}{a_t},
\textcolor{recentcolor}{o_t},
\textcolor{recentcolor}{\iota_t})  \\
&\sim \pi_\theta(\cdot \mid I_t,\mathcal{S}_t),
\end{aligned}
\end{equation}
where $\tau_t$ is reasoning, \textcolor{histcolor}{$\phi_t$} is a folding directive, \textcolor{memstatecolor}{$a_t$} is a UI or memory action, \textcolor{recentcolor}{$o_t$} is the UI observation, and \textcolor{recentcolor}{$\iota_t$} is the action intent. Unlike ReAct-style prompting, which appends a growing task-progress string, \conact partitions context into three fields and updates them through model-emitted context actions.

\paragraph{Step output protocol.}
At each step, the model emits a \textbf{5-part structured output}. We build on the Qwen3-VL function-calling prompt format~\cite{bai2025qwen3}, which asks for a concise thought and a JSON tool call enclosed by \texttt{<tool\_call>} tags. \conact adds context-management fields for folding and self-description, and extends the executable tool-call action space with memory operations. The fields are colored by their roles, consistent with Figure~\ref{fig:framework}:

\begin{Verbatim}[commandchars=\\\{\},fontsize=\footnotesize]
\textcolor{black}{<thinking>} reasoning \textcolor{black}{</thinking>}
\textcolor{histcolor}{<folding>} \{range, summary\} \textcolor{histcolor}{</folding>}
\textcolor{memstatecolor}{<tool_call>} UI / memory action \textcolor{memstatecolor}{</tool_call>}
\textcolor{recentcolor}{<ui_observation>} screen facts \textcolor{recentcolor}{</ui_observation>}
\textcolor{recentcolor}{<action_intent>} next-step plan \textcolor{recentcolor}{</action_intent>}
\end{Verbatim}

The \toolfield{<tool\_call>} field executes exactly one action per step from the extended action space
\begin{equation}
\label{eq:action-space}
\begin{aligned}
\mathcal{A} &= \mathcal{A}_{ui}\cup\mathcal{A}_{mem},\\
\mathcal{A}_{mem} &= \{\memaction{add},\memaction{update},\memaction{delete}\}.
\end{aligned}
\end{equation}
Here, $\mathcal{A}_{ui}$ contains UI actions such as clicking, typing, swiping, waiting, and terminating. We use \memaction{add}, \memaction{update}, and \memaction{delete} as shorthand for \memaction{memory\_add}, \memaction{memory\_update}, and \memaction{memory\_delete}, which update \textcolor{memstatecolor}{$M_t$}. Other context fields do not act on the environment; they specify how the working context is updated for future steps.

\paragraph{\symfold History folding (over \textcolor{histcolor}{$H_t$}).}
\label{sec:method-fold}
From step 2 onward, the agent emits a mandatory \foldop{<folding>} block. Formally, the folding directive is \textcolor{histcolor}{$\phi_t=([s_t,t],z_t)$}, where $[s_t,t]$ is the history span to compress and $z_t$ is the generated summary. \emph{Step-level distillation} corresponds to $s_t=t$ and keeps the latest step as a compact record, while \emph{span-level abstraction} uses $s_t<t$ to summarize a completed sub-task into one reusable record. The folded history is updated as
\begin{equation}
\label{eq:history-update}
\textcolor{histcolor}{H_{t+1}}
=
\mathrm{Fold}(
\textcolor{histcolor}{H_t},
\textcolor{histcolor}{\phi_t}
),
\end{equation}
which replaces uncontrolled \texttt{<conclusion>} accumulation with model-controlled folding, pushing context growth from linear accumulation toward sub-linear growth.

\paragraph{\symmem UI memory actions (over \textcolor{memstatecolor}{$M_t$}).}
\label{sec:method-mem}
Each memory item is a structured triple $m=(\mathrm{id},d,c)$, where $\mathrm{id}$ is a unique identifier, $d$ is a short description, and $c$ is the complete content to preserve. Let $\mathcal{A}_{mem}=\{a^{+},a^{\circ},a^{-}\}$ denote \memaction{memory\_add}, \memaction{memory\_update}, and \memaction{memory\_delete}. A memory action induces
\begin{equation}
\label{eq:memory-update}
\textcolor{memstatecolor}{M_{t+1}}=
\begin{cases}
\mathrm{Add}(\textcolor{memstatecolor}{M_t},m),
& a_t=a^{+},\\
\mathrm{Update}(\textcolor{memstatecolor}{M_t},\mathrm{id},c),
& a_t=a^{\circ},\\
\mathrm{Delete}(\textcolor{memstatecolor}{M_t},\mathrm{id}),
& a_t=a^{-},\\
\textcolor{memstatecolor}{M_t},
& a_t\in\mathcal{A}_{ui}.
\end{cases}
\end{equation}
Memory writes store complete task-relevant information rather than references or lossy summaries, e.g., a full copied price or phone number instead of ``the value seen earlier.'' This is crucial when prices, codes, contact information, or copied text must survive screen changes, long delays, and app transitions.

\paragraph{\symstep Self-describing step output (fills \textcolor{recentcolor}{$L_t$}).}
\label{sec:method-step}
The two self-describing fields make each step reusable by future context actions. \recentfield{<ui\_observation>} provides a grounded screen description with exact visible text, numbers, names, and task-relevant UI facts. \recentfield{<action\_intent>} states what the current tool call is intended to accomplish. Together with the executed action and tool result, they form
\begin{equation}
\label{eq:recent-update}
\textcolor{recentcolor}{L_{t+1}}
=
(
\textcolor{recentcolor}{o_t},
\textcolor{recentcolor}{\iota_t},
\textcolor{memstatecolor}{a_t},
r_t
).
\end{equation}
This record supplies grounded content for memory writes into \textcolor{memstatecolor}{$M_t$} and future folding into \textcolor{histcolor}{$H_t$}. Without these fields, folding and memory actions must infer step meaning only from raw \texttt{<thinking>} and \texttt{<tool\_call>} text, which we find insufficient when context actions are layered onto the ReAct 3-part protocol (\S\ref{sec:ablation}).

\paragraph{Step execution loop.}
At step $t$, the MLLM $\pi_\theta$ takes $(I_t,\mathcal{S}_t)$ and emits $y_t=(\tau_t,\phi_t,a_t,o_t,\iota_t)$. The environment/tool result is
\begin{equation}
\label{eq:tool-result}
(r_t,I_{t+1})=
\begin{cases}
\mathrm{Env}(I_t,a_t), & a_t\in\mathcal{A}_{ui},\\
(\mathrm{ok},I_t), & a_t\in\mathcal{A}_{mem},
\end{cases}
\end{equation}
where UI actions change the screen and memory actions update only the structured context. Combining Eq.~\ref{eq:history-update}, Eq.~\ref{eq:memory-update}, and Eq.~\ref{eq:recent-update}, the complete state transition is
\begin{equation}
\label{eq:state-transition}
\begin{aligned}
\mathcal{S}_{t+1}
&=
\mathcal{T}(\mathcal{S}_t,y_t,r_t)\\
&=
(G,\textcolor{histcolor}{H_{t+1}},
\textcolor{memstatecolor}{M_{t+1}},
\textcolor{recentcolor}{L_{t+1}}).
\end{aligned}
\end{equation}
The next state $\mathcal{S}_{t+1}$ is fed into step $t{+}1$. Because folding, memory, and action intent are predicted by the same multimodal policy in one forward pass, compression and memorization inherit the agent's task-level reasoning rather than being delegated to a separate summarizer, retriever, or memory agent.

\section{MemGUI-3K Dataset}
\label{sec:dataset}
\vspace{-0.25em}

\subsection{Motivation: Context Management Must Be Learned}
\label{sec:dataset-motivation}

\begin{table}[!htbp]
  \centering
  \footnotesize
  \renewcommand{\arraystretch}{1.06}
  \setlength{\tabcolsep}{3.6pt}
  \begin{adjustbox}{max width=0.88\linewidth}
  \begin{tabular}{@{}l | cc | cc | cc | cc@{}}
    \toprule
    & \multicolumn{2}{c|}{\textbf{P@1 (\%)}}
    & \multicolumn{2}{c|}{\textbf{P@2 (\%)}}
    & \multicolumn{2}{c|}{\textbf{P@3 (\%)}}
    & \multicolumn{2}{c}{\textbf{IRR (\%)}} \\
    \textbf{Model Scale}
      & \texttt{Base} & \conact
      & \texttt{Base} & \conact
      & \texttt{Base} & \conact
      & \texttt{Base} & \conact \\
    \midrule
    Qwen3-VL-2B
      & 10.0 & \drop{5.0}    & 12.5 & \drop{5.0}    & 17.5 & \drop{10.0}
      & 10.8 & \drop{1.6}  \\
    Qwen3-VL-4B
      & 20.0 & \drop{12.5}   & 25.0 & \drop{15.0}   & 27.5 & \drop{15.0}
      & 19.0 & \drop{10.7} \\
    Qwen3-VL-8B
      & 12.5 & \drop{7.5}    & 17.5 & \drop{10.0}   & 22.5 & \drop{15.0}
      & 15.0 & \drop{10.7} \\
    \midrule
    Qwen3-VL-235B-Instruct
      & 23.4 & \drop{19.5}   & 36.7 & \drop{23.4}   & 36.7 & \drop{28.9}
      & 29.2 & \rise{32.5} \\
    \midrule
    \rowcolor{oursrowcolor} \textbf{Qwen3-VL-235B-Thinking}
      & 5.0  & \rise{\textbf{40.0}}  & 20.0 & \rise{\textbf{52.5}}  & 27.5 & \rise{\textbf{62.5}}
      & 19.5 & \rise{\textbf{51.0}} \\
    \bottomrule
  \end{tabular}%
  \end{adjustbox}
  \caption{Zero-shot \conact across \llmname{Qwen3-VL} sizes on MemGUI-Bench-40. \textbf{Base}: ReAct-style 3-part prompt; \textbf{\conact}: 5-part context-as-action prompt. \rise{Green}/\drop{red} denote improvement/regression. Only \llmname{235B-Thinking} benefits zero-shot, motivating MemGUI-3K supervision.}
  \label{tab:ablation-model-size}
\end{table}

\conact provides an end-to-end interface for folding history, writing UI memory, and describing recent steps, but using it well still requires policy decisions. We therefore ask whether the protocol alone is sufficient across model scales. Table~\ref{tab:ablation-model-size} applies the same 5-part \conact protocol to \llmname{Qwen3-VL} backbones of different sizes on MemGUI-Bench-40. Gains emerge only on the strongest backbone we test, \llmname{Qwen3-VL-235B-Thinking}; on smaller open models and \llmname{Qwen3-VL-235B-Instruct}, the added protocol often reduces performance. This suggests that proactive context management is not merely a prompt-format change: models must learn when to fold history, when to write UI facts into memory, and how to produce reusable step descriptions. We therefore construct \textbf{MemGUI-3K}, a supervised dataset with full \conact annotations, to support training and analysis across model scales.

\subsection{Construction Pipeline}
\label{sec:dataset-construction}

\begin{figure}[h]
  \centering
  \includegraphics[width=0.86\linewidth]{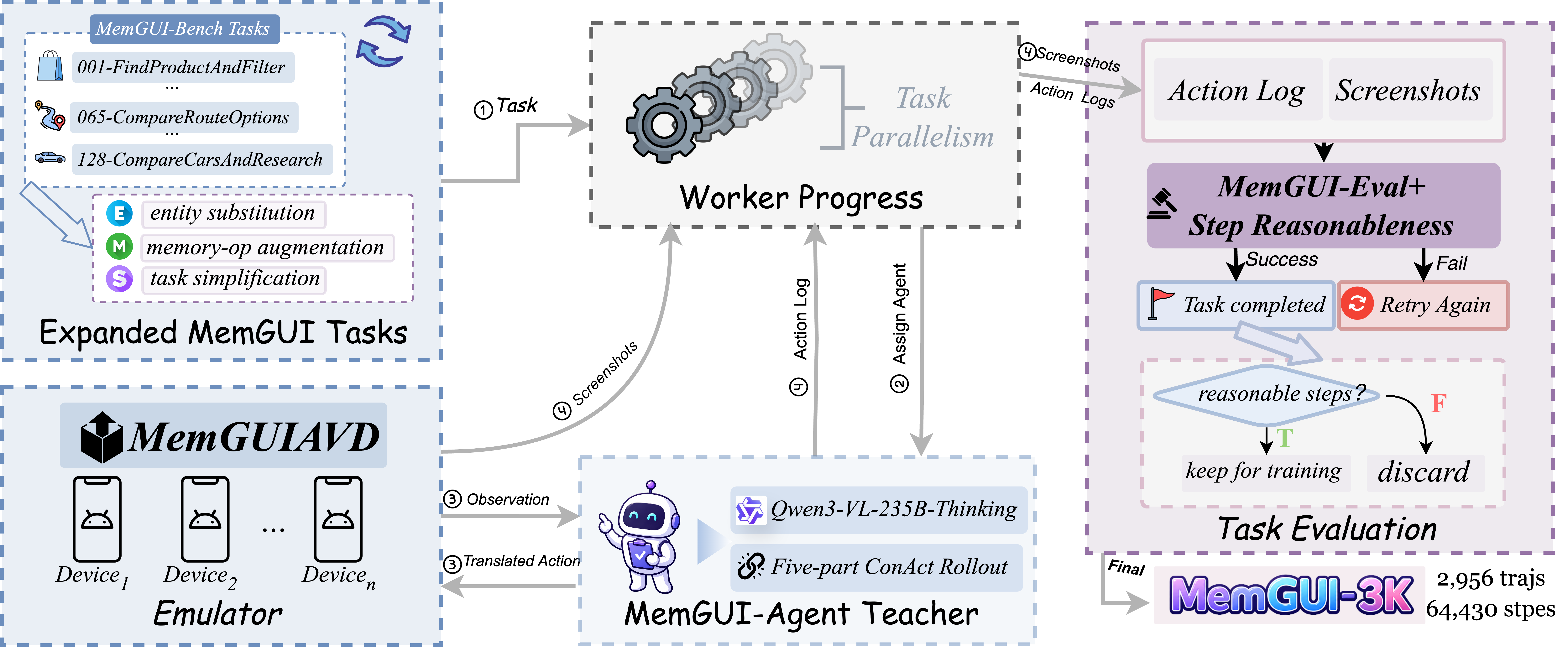}
  \caption{MemGUI-3K data collection and filtering pipeline. Expanded tasks are rolled out by a \llmname{Qwen3-VL-235B-Thinking} teacher, evaluated by MemGUI-Eval, filtered at trajectory and step levels, and converted into SFT samples.}
  \label{fig:data-pipeline}
\end{figure}

\textbf{Expanding seed tasks.}
MemGUI-3K is built from the 128 seed tasks of MemGUI-Bench~\cite{liu2026memgui}. We expand them with three complementary strategies: \emph{entity substitution}, which preserves task structure while replacing task-specific entities; \emph{memory-operation augmentation}, which increases coverage of \memaction{memory\_update} and \memaction{memory\_delete}; and \emph{task simplification}, which decomposes complex multi-app tasks into shorter single-objective variants. This produces a 7{,}303-task pool, from which 5{,}293 tasks enter rollout. Detailed expansion statistics are provided in Appendix~\ref{sec:appendix-task-expansion}.

\textbf{Collecting teacher rollouts.}
Each task is executed once in the snapshot-based Android environment of MemGUI-Bench using \llmname{Qwen3-VL-235B-Thinking} as the teacher with the full 5-part \conact protocol. Rollouts terminate when the agent emits \texttt{terminate} or reaches the task-specific budget $2.5g+1$, where $g$ is the task's golden-step count. We evaluate trajectories with the MemGUI-Eval progressive scrutiny pipeline~\cite{liu2026memgui} and additionally annotate every step as \emph{reasonable} or \emph{unreasonable}. This filtering removes recovery loops, repeated errors, and counterproductive actions from SFT targets while preserving successful context-management behavior. Figure~\ref{fig:data-pipeline} visualizes the workflow; details and an annotated example are in Appendices~\ref{sec:appendix-rollout-filtering}--\ref{sec:appendix-training-data-example}.

\textbf{Filtering into training samples.}
After trajectory filtering and a 90/10 split, MemGUI-3K contains 2{,}956 successful trajectories across 26 apps, with verified zero overlap against the 128 MemGUI-Bench evaluation tasks. Extracting reasonable steps yields 64{,}430 SFT samples, including 57{,}951 train and 6{,}479 test samples. Each sample consists of the shared system prompt, the user message with structured context state and screenshot, and the gold 5-part \conact response. We train \textbf{MemGUI-8B-SFT} from \llmname{Qwen3-VL-8B-Instruct} using LoRA SFT with ms-swift~\cite{zhao2024swift}; dataset format and training details are provided in Appendix~\ref{sec:appendix-dataset-format} and Appendix~\ref{sec:appendix-training-setup}.

\subsection{Statistics}
\label{sec:dataset-statistics}

\begin{figure}[h]
  \centering
  \includefig[width=0.88\linewidth]{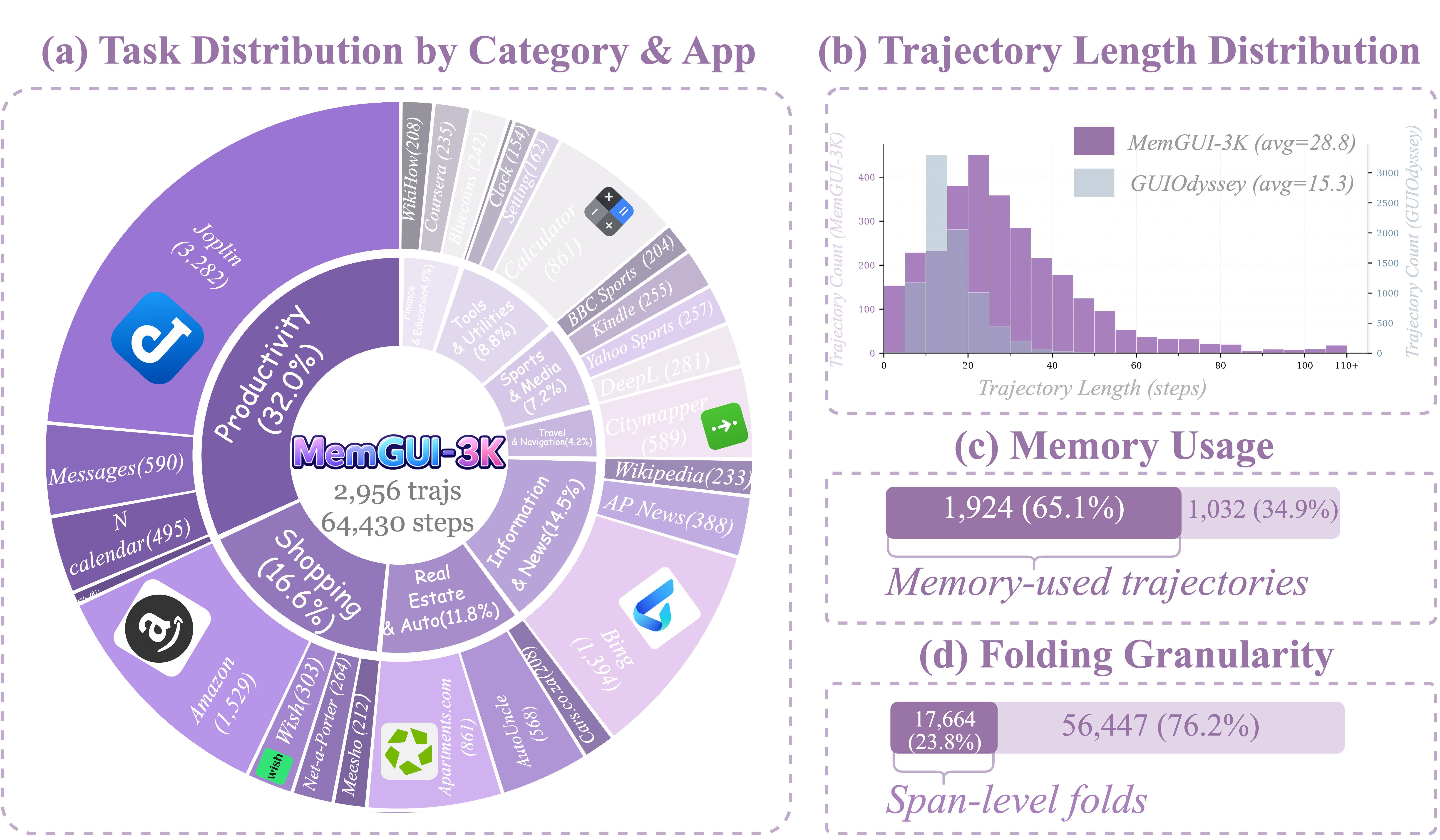}
  \caption{MemGUI-3K statistics. \textbf{(a)}~Task distribution over 26 apps and 7 categories. \textbf{(b)}~Trajectory length: avg.\ 28.8 vs.\ 15.3 for GUIOdyssey~\cite{lu2025guiodyssey} ($1.9\times$). \textbf{(c)}~Memory-action usage. \textbf{(d)}~Folding granularity, separating span-level abstraction from step-level distillation.}
  \label{fig:dataset-stats}
\end{figure}

Figure~\ref{fig:dataset-stats} summarizes MemGUI-3K. The dataset covers 26 Android apps across 7 functional categories, with productivity, shopping, and information~\&~news being the most frequent (Figure~\ref{fig:dataset-stats}(a)). Trajectories average 28.8 steps with a median of 25, about $1.9\times$ longer than GUIOdyssey~\cite{lu2025guiodyssey} (avg.\ 15.3), the previously longest mobile GUI SFT dataset (Figure~\ref{fig:dataset-stats}(b)). Step-level reasonableness analysis marks 75.7\% of steps as reasonable, yielding 21.8 trainable steps per trajectory on average. Memory actions appear in 65.1\% of trajectories (Figure~\ref{fig:dataset-stats}(c)). The teacher also performs nontrivial history abstraction: 23.8\% of folds are span-level summaries, with an average span of 6.25 steps and 88.7\% of trajectories containing at least one span-level fold (Figure~\ref{fig:dataset-stats}(d)). Additional statistics on splits, task sources, trajectory lengths, and folding granularity are reported in Appendix~\ref{sec:appendix-dataset-statistics}. These statistics show that MemGUI-3K supervises UI actions, memory writing, history compression, and reusable step descriptions.

\begin{table*}[t]
  \centering
  \footnotesize
  \renewcommand{\arraystretch}{1.06}
  \setlength{\tabcolsep}{3.2pt}
  \begin{adjustbox}{max width=\textwidth}
  \begin{tabular}{@{}ll | ccc | ccc | ccc | ccc@{}}
    \toprule
    \multirow{2}{*}{\textbf{Agent}} & \multirow{2}{*}{\textbf{Memory Type}} &
      \multicolumn{3}{c|}{\textbf{Easy}} &
      \multicolumn{3}{c|}{\textbf{Medium}} &
      \multicolumn{3}{c|}{\textbf{Hard}} &
      \multicolumn{3}{c}{\textbf{Overall}} \\
    \cmidrule(lr){3-5} \cmidrule(lr){6-8} \cmidrule(lr){9-11} \cmidrule(l){12-14}
    & & \texttt{P@1} & \texttt{P@3} & \texttt{IRR} & \texttt{P@1} & \texttt{P@3} & \texttt{IRR} & \texttt{P@1} & \texttt{P@3} & \texttt{IRR} & \texttt{P@1} & \texttt{P@3} & \texttt{IRR} \\
    \midrule
    \multicolumn{14}{l}{\textit{\textbf{Agentic Frameworks}} \textit{(w/ Gemini-2.5-Pro backbone)}} \\
    \midrule
    Agent-S2~\cite{agashe2025agent} & \textit{Memory Agent} & \textbf{41.7} & \underline{64.6} & 38.7 & 19.0 & 42.9 & 42.5 & 18.4 & 36.8 & \underline{36.9} & 27.3 & \underline{49.2} & \underline{39.5} \\
    M3A~\cite{rawles2025androidworld} & \textit{Memory Agent} & \underline{39.6} & 47.9 & 31.7 & \textbf{35.7} & \underline{50.0} & \underline{48.7} & \underline{21.1} & \underline{44.7} & 36.4 & \underline{32.8} & 47.7 & 39.3 \\
    T3A~\cite{rawles2025androidworld} & \textit{Memory Agent} & 31.2 & 45.8 & -- & 16.7 & 45.2 & -- & 18.4 & 34.2 & -- & 22.7 & 42.2 & 29.6 \\
    Mobile-Agent-E~\cite{wang2025mobileagentselfevolving} & \textit{Memory Agent} & 12.5 & 22.9 & 0.9 & 2.4 & 2.4 & 4.4 & 0.0 & 2.6 & 1.8 & 5.5 & 10.2 & 2.4 \\
    Mobile-Agent-V2~\cite{wang2024mobile} & \textit{Memory Agent} & 8.3 & 10.4 & 0.0 & 0.0 & 0.0 & 0.0 & 0.0 & 0.0 & 0.0 & 3.1 & 3.9 & 0.0 \\
    SeeAct~\cite{zheng2024seeact} & \textit{Rule-based} & 6.2 & 12.5 & 0.0 & 0.0 & 2.4 & 0.5 & 0.0 & 0.0 & 0.0 & 2.3 & 5.5 & 0.2 \\
    AppAgent~\cite{zhang2025appagent} & \textit{Action-Thought} & 8.3 & 22.9 & 4.6 & 0.0 & 2.4 & 0.0 & 0.0 & 0.0 & 0.0 & 3.1 & 9.4 & 1.5 \\
    \midrule
    \multicolumn{14}{l}{\textit{\textbf{End-to-End Models}}} \\
    \midrule
    Qwen3-VL-235B-Thinking~\cite{bai2025qwen3} & \textit{Action-Thought} & 18.8 & 47.9 & 21.8 & \underline{28.6} & 47.6 & 38.8 & \underline{26.3} & \underline{44.7} & 28.2 & 24.2 & 46.9 & 30.0 \\
    Qwen3-VL-235B-Instruct~\cite{bai2025qwen3} & \textit{Action-Thought} & 27.1 & 39.6 & 26.5 & 21.4 & 40.5 & 28.8 & 21.1 & 28.9 & 32.2 & 23.4 & 36.7 & 29.2 \\
    Qwen3-VL-32B-Instruct~\cite{bai2025qwen3} & \textit{Action-Thought} & 35.4 & 47.9 & \underline{39.7} & 14.3 & 23.8 & 22.4 & 7.9 & 21.1 & 13.8 & 20.3 & 32.0 & 25.0 \\
    GUI-Owl-1.5-32B-Instruct~\cite{xu2026mobile} & \textit{Action-Thought} & 22.9 & 37.5 & 18.4 & 7.1 & 14.3 & 21.3 & 0.0 & 5.3 & 15.3 & 10.9 & 20.3 & 18.4 \\
    Qwen3-VL-8B-Instruct~\cite{bai2025qwen3} & \textit{Action-Thought} & 18.8 & 35.4 & 20.1 & 4.8 & 14.3 & 16.0 & 2.6 & 7.9 & 9.5 & 9.4 & 20.3 & 15.1 \\
    GUI-Owl-1.5-8B-Instruct~\cite{xu2026mobile} & \textit{Action-Thought} & 22.9 & 31.2 & 15.7 & 2.4 & 2.4 & 16.5 & 7.9 & 10.5 & 14.2 & 11.7 & 15.6 & 15.5 \\
    GUI-Owl-7B~\cite{ye2025mobile} & \textit{Action-Thought} & 14.6 & 22.9 & 9.9 & 0.0 & 2.4 & 5.0 & 2.6 & 2.6 & 2.6 & 6.2 & 10.2 & 5.7 \\
    UI-TARS-1.5-7B~\cite{qin2025ui} & \textit{Multi-turn} & 8.3 & 16.7 & 9.3 & 0.0 & 0.0 & 1.7 & 0.0 & 0.0 & 0.9 & 3.1 & 6.2 & 3.8 \\
    UI-Venus-7B~\cite{gu2025ui-venus} & \textit{Action-Thought} & 14.6 & 20.8 & 6.9 & 0.0 & 0.0 & 1.2 & 0.0 & 0.0 & 0.0 & 5.5 & 7.8 & 2.6 \\
    CogAgent~\cite{hong2024cogagent} & \textit{No History} & 0.0 & 0.0 & 0.0 & 0.0 & 0.0 & 0.0 & 0.0 & 0.0 & 0.0 & 0.0 & 0.0 & 0.0 \\
    \midrule
    \addlinespace[2pt]
    \rowcolor{oursrowcolor} \textbf{\ourmethod-235B} \textit{(Ours, zero-shot)} & \conact & \textbf{41.7} & \textbf{68.8} & \textbf{40.8} & \textbf{35.7} & \textbf{61.9} & \textbf{52.3} & \textbf{34.2} & \textbf{55.3} & \textbf{46.5} & \textbf{37.5} & \textbf{62.5} & \textbf{46.8} \\[-2pt]
    \multicolumn{2}{l|}{\quad\textit{\footnotesize $\Delta$ vs.\ Qwen3-VL-235B-Thinking}}
      & {\footnotesize\rise{+22.9}} & {\footnotesize\rise{+20.9}} & {\footnotesize\rise{+19.0}}
      & {\footnotesize\rise{+7.1}} & {\footnotesize\rise{+14.3}} & {\footnotesize\rise{+13.5}}
      & {\footnotesize\rise{+7.9}} & {\footnotesize\rise{+10.6}} & {\footnotesize\rise{+18.3}}
      & {\footnotesize\rise{+13.3}} & {\footnotesize\rise{+15.6}} & {\footnotesize\rise{+16.8}} \\
    \addlinespace[2pt]
    \rowcolor{oursrowcolor} \textbf{MemGUI-8B-SFT} \textit{(Ours)} & \conact & 25.0 & 39.6 & 20.5 & 23.8 & 33.3 & 36.5 & 21.1 & 34.2 & 32.6 & 23.4 & 35.9 & 30.2 \\[-2pt]
    \multicolumn{2}{l|}{\quad\textit{\footnotesize $\Delta$ vs.\ Qwen3-VL-8B-Instruct}}
      & {\footnotesize\rise{+6.2}} & {\footnotesize\rise{+4.2}} & {\footnotesize\rise{+0.4}}
      & {\footnotesize\rise{+19.0}} & {\footnotesize\rise{+19.0}} & {\footnotesize\rise{+20.5}}
      & {\footnotesize\rise{+18.5}} & {\footnotesize\rise{+26.3}} & {\footnotesize\rise{+23.1}}
      & {\footnotesize\rise{+14.0}} & {\footnotesize\rise{+15.6}} & {\footnotesize\rise{+15.1}} \\
    \bottomrule
  \end{tabular}%
  \end{adjustbox}
  \vspace{-2mm}
  \caption{Performance breakdown on MemGUI-Bench by difficulty: Easy (1--20 steps), Medium (21--40), Hard (41+). Both of our agents use a \llmname{Qwen3-VL} backbone with \conact: \ourmethod-235B applies \conact zero-shot to \llmname{Qwen3-VL-235B-Thinking} (weights unchanged); MemGUI-8B-SFT is fine-tuned on MemGUI-3K from \llmname{Qwen3-VL-8B-Instruct}. \textbf{Bold}/\underline{underline} = best/second-best (ties share bold); ``--'' = unavailable. $\Delta$ rows show gains over the corresponding ReAct-style baseline.}
  \label{tab:main-results}
\end{table*}

\section{Experiments}
\label{sec:experiments}
\setcounter{insight}{0}
\vspace{-0.25em}

\paragraph{Evaluation setup.}
We evaluate on MemGUI-Bench~\cite{liu2026memgui}, MobileWorld GUI-Only~\cite{kong2025mobileworld}, and MemGUI-Bench-40 for ablations and failure analysis. We report Pass@$k$ ($k{\in}\{1,3\}$), IRR, and MTPR on MemGUI-Bench, and single-attempt success rate on MobileWorld. Baselines are listed in Tables~\ref{tab:main-results} and~\ref{tab:mobileworld-results}; our two reported agents are \textbf{\ourmethod-235B}, a zero-shot \conact agent using \llmname{Qwen3-VL-235B-Thinking}, and \textbf{MemGUI-8B-SFT}, an SFT \llmname{Qwen3-VL-8B-Instruct} agent trained on MemGUI-3K. Unless otherwise stated, ablations, case studies, and failure analysis use zero-shot \ourmethod-235B. Benchmark details and prompts are in Appendices~\ref{sec:appendix-eval-protocol} and~\ref{sec:appendix-prompts}.

\subsection{Main Results}
\label{sec:main-results}
\vspace{-0.25em}

\paragraph{Zero-shot \conact sets a new SOTA on MemGUI-Bench.}
On MemGUI-Bench (Table~\ref{tab:main-results}), \ourmethod-235B reaches 37.5\% Pass@1 and 62.5\% Pass@3, surpassing the best agentic framework, M3A on \llmname{Gemini-2.5-Pro} (32.8\% / 47.7\%), and improving over the same backbone by $+$13.3 Pass@1 and $+$16.8 IRR. On MobileWorld (Table~\ref{tab:mobileworld-results}), \ourmethod-235B achieves 29.1\% SR, $+$14.6 over the \llmname{Qwen3-VL-235B-Thinking} baseline, showing that zero-shot \conact also transfers to a different environment and app set. As Figure~\ref{fig:main-performance}(a) shows, these gains are not explained by larger prompts alone: \conact keeps prompt growth substantially flatter than the ReAct baseline over 150 steps.

\begin{table}[!htbp]
  \centering
  \footnotesize
  \renewcommand{\arraystretch}{1.06}
  \setlength{\tabcolsep}{4pt}
  \begin{adjustbox}{max width=0.92\linewidth}
  \begin{tabular}{@{}l l c@{}}
    \toprule
    \textbf{Agent} & \textbf{Memory Type} & \textbf{GUI-Only SR (\%)} \\
    \midrule
    \multicolumn{3}{l}{\textit{\textbf{Open-Weight Models}}} \\
    \midrule
    GUI-Owl-1.5-32B-Instruct~\cite{xu2026mobile} & \textit{Action-Thought} & 43.9 \\
    MAI-UI-235B-A22B~\cite{zhou2025mai} & \textit{Action-Thought} & 39.7 \\
    GUI-Owl-1.5-8B-Instruct~\cite{xu2026mobile} & \textit{Action-Thought} & 38.2 \\
    MAI-UI-32B~\cite{zhou2025mai} & \textit{Action-Thought} & 36.2 \\
    GUI-Owl-1.5-2B-Instruct~\cite{xu2026mobile} & \textit{Action-Thought} & 32.2 \\
    MAI-UI-8B~\cite{zhou2025mai} & \textit{Action-Thought} & 27.5 \\
    Qwen3-VL-235B-Thinking~\cite{bai2025qwen3} & \textit{Action-Thought} & 14.5 \\
    Qwen3-VL-235B-Instruct~\cite{bai2025qwen3} & \textit{Action-Thought} & 12.8 \\
    Qwen3-VL-32B-Instruct~\cite{bai2025qwen3} & \textit{Action-Thought} & 11.9 \\
    Qwen3-VL-8B-Instruct~\cite{bai2025qwen3} & \textit{Action-Thought} & 9.4 \\
    \midrule
    \multicolumn{3}{l}{\textit{\textbf{Open-Data Models}}} \\
    \midrule
    OpenMobile-8B~\cite{OpenMobile2025} & \textit{Action-Thought} & 17.7 \\
    ClawGUI-2B~\cite{tang2026clawgui} & \textit{Action-Thought} & 17.1 \\
    OpenMobile-7B~\cite{OpenMobile2025} & \textit{Action-Thought} & 14.8 \\
    ScaleCUA-7B~\cite{ScaleCUA2025} & \textit{Action-Thought} & 7.7 \\
    \midrule
    \addlinespace[2pt]
    \rowcolor{oursrowcolor} \textbf{\ourmethod-235B} \textit{(Ours, zero-shot)} & \conact & \textbf{29.1} \\
    \multicolumn{2}{l}{\quad\textit{\footnotesize $\Delta$ vs.\ Qwen3-VL-235B-Thinking}} & {\footnotesize\rise{+14.6}} \\
    \addlinespace[2pt]
    \rowcolor{oursrowcolor} \textbf{MemGUI-8B-SFT} \textit{(Ours)} & \conact & \underline{17.9} \\
    \multicolumn{2}{l}{\quad\textit{\footnotesize $\Delta$ vs.\ Qwen3-VL-8B-Instruct}} & {\footnotesize\rise{+8.5}} \\
    \bottomrule
  \end{tabular}%
  \end{adjustbox}
  \vspace{-2mm}
  \caption{Pass@1 success rate (\%) on MobileWorld GUI-Only (117 tasks). \textbf{Bold}/\underline{underline} = best/second-best among open-data models. $\Delta$ rows show gains over the corresponding \llmname{Qwen3-VL} backbone.}
  \label{tab:mobileworld-results}
\end{table}

\paragraph{MemGUI-3K enables an 8B agent and transfers beyond its source benchmark.}
MemGUI-8B-SFT reaches 23.4\% Pass@1 on MemGUI-Bench versus 9.4\% for the \llmname{Qwen3-VL-8B-Instruct} baseline, with gains concentrated on harder tasks (Medium $+$19.0, Hard $+$18.5 vs.\ Easy $+$6.2). This difficulty trend indicates that MemGUI-3K does not merely teach a new output format; it improves behavior in the regime where long-horizon context drift is most severe. On MobileWorld, where the environment, apps, and evaluation protocol all differ, MemGUI-8B-SFT still reaches 17.9\% SR ($+$8.5 over backbone) and ranks among the best open-data 8B models. The learned context-management skills therefore transfer beyond the source benchmark rather than only fitting MemGUI-Bench.

\vspace{-0.25em}
\subsection{Offline Skill Analysis}
\label{sec:offline-skill-analysis}
\vspace{-0.55em}

\begin{table}[!htbp]
  \centering
  \footnotesize
  \renewcommand{\arraystretch}{1.06}
  \setlength{\tabcolsep}{4pt}
  \begin{adjustbox}{max width=0.86\linewidth}
  \begin{tabular}{@{}l | c | c | cc | c@{}}
    \toprule
    & \textbf{UI Action} & \textbf{Memory} & \multicolumn{2}{c|}{\textbf{Folding}} & \textbf{Format} \\
    \textbf{Agent} & \texttt{Match} & \texttt{Trigger F1} & \texttt{Deep Ratio} & \texttt{Range Acc.} & \texttt{Tags} \\
    \midrule
    \texttt{8B-Inst.} \textit{(ZS)}
      & 29.2 & 19.9 & 8.8 & 45.2 & 94.9 \\
    \addlinespace[2pt]
    \rowcolor{oursrowcolor} \textbf{MemGUI-8B-SFT} \textit{(Ours)}
      & \textbf{36.3} & \textbf{48.0} & \textbf{26.1} & \textbf{58.9} & \textbf{99.9} \\[-2pt]
    \multicolumn{1}{l|}{\quad\textit{\footnotesize $\Delta$ vs.\ 8B-Inst.}}
      & {\footnotesize\rise{+7.1}} & {\footnotesize\rise{+28.1}} & {\footnotesize\rise{+17.3}} & {\footnotesize\rise{+13.7}} & {\footnotesize\rise{+5.0}} \\
    \addlinespace[2pt]
    \ourmethod-235B \textit{(ZS)}
      & \underline{33.9} & \underline{34.3} & \underline{18.7} & \underline{56.7} & \underline{99.2} \\
    \bottomrule
  \end{tabular}%
  \end{adjustbox}
  \caption{Core gold-context step-level offline metrics on the MemGUI-3K test set. Each agent receives the gold SFT input for a single step and is compared with the gold assistant response. All scores are percentages. \texttt{8B-Inst.} is \llmname{Qwen3-VL-8B-Instruct}; ZS denotes zero-shot. \textbf{Bold}/\underline{underline} = best/second-best; $\Delta$ shows gains over the 8B backbone. Detailed metrics are in Table~\ref{tab:offline-eval-detailed}.}
  \label{tab:offline-eval}
\end{table}

\vspace{-0.15em}

Online benchmarks measure end-to-end success but obscure subskills. We conduct step-level offline evaluation on the MemGUI-3K test set (295 trajectories, 6{,}479 steps): for each reasonable step, the evaluator feeds the same SFT input, consisting of the system prompt, structured context state, and screenshot, and compares the generated response with the gold response. This isolates whether the model learns action, memory, folding, and formatting decisions. Table~\ref{tab:offline-eval} reports core metrics, with details in Table~\ref{tab:offline-eval-detailed}. \emph{UI actions} improve moderately: match accuracy rises from 29.2\% to 36.3\%, with larger gains on click and type. \emph{Memory timing} improves most: trigger F1 rises from 19.9\% to 48.0\%, exceeding zero-shot \ourmethod-235B (34.3\%). \emph{Deep folding} also improves: MemGUI-8B-SFT reaches 26.1\% deep ratio and 58.9\% deep range accuracy, compared with 8.8\% deep ratio for zero-shot 8B and 23.2\% gold. \emph{Format compliance} reaches 99.9\%, confirming reliable structured generation. Metric definitions are in Appendix~\ref{sec:appendix-offline-skill}.

\insight{MemGUI-3K teaches context control decisions, not just formatting: the largest gain comes from promoting transient observations to durable memory.}

\subsection{Ablation Study}
\label{sec:ablation}
\vspace{-0.35em}

\begin{table}[!htbp]
  \centering
  \footnotesize
  \renewcommand{\arraystretch}{1.06}
  \setlength{\tabcolsep}{4pt}
  \begin{adjustbox}{max width=0.88\linewidth}
  \begin{tabular}{@{}l | ccc | cc@{}}
    \toprule
    & \multicolumn{3}{c|}{\textbf{Success Rate}} & \multicolumn{2}{c}{\textbf{Memory Metrics}} \\
    \textbf{Variant} & \texttt{P@1} & \texttt{P@2} & \texttt{P@3} & \texttt{MTPR}$^\dagger$ & \texttt{IRR}$^\ddagger$ \\
    \midrule
    ReAct baseline
      & 5.0  & 20.0 & 27.5
      & 0.143 & 19.5 \\
    \midrule
    \quad + UI memory actions
      & 17.5 & 35.0 & 42.5
      & \underline{0.357} & \underline{39.1} \\
    \quad + history folding
      & 22.5 & 25.0 & 32.5
      & 0.179 & 25.7 \\
    \quad + self-describing step
      & \underline{25.0} & \underline{40.0} & \underline{45.0}
      & 0.214 & 33.1 \\
    \midrule
    \addlinespace[2pt]
    \rowcolor{oursrowcolor} \textbf{Full \conact} \textit{(Ours)}
      & \textbf{40.0} & \textbf{52.5} & \textbf{62.5}
      & \textbf{0.429} & \textbf{51.0} \\
    \multicolumn{1}{l}{\quad\textit{\footnotesize $\Delta$ vs.\ ReAct baseline}}
      & {\footnotesize\rise{+35.0}} & {\footnotesize\rise{+32.5}} & {\footnotesize\rise{+35.0}}
      & {\footnotesize\rise{+0.286}} & {\footnotesize\rise{+31.5}} \\
    \bottomrule
  \end{tabular}%
  \end{adjustbox}
  \caption{Component ablation of \conact on MemGUI-Bench-40 (\llmname{Qwen3-VL-235B-Thinking}, zero-shot). Each middle row adds one mechanism to the ReAct baseline; the last row enables all three. \textbf{Bold}/\underline{underline} = best/second-best (among single-component variants for underline). $^\dagger$MTPR: Memory-Task Proficiency Ratio. $^\ddagger$IRR: Information Retention Rate (\%).}
  \label{tab:ablation-study}
\end{table}

\vspace{-0.15em}

We ablate the three components of \conact on MemGUI-Bench-40 with zero-shot \llmname{Qwen3-VL-235B-Thinking} (Table~\ref{tab:ablation-study}). Each targets a different failure mode, and none suffices alone. UI memory actions triple Pass@1 from 5.0\% to 17.5\%, showing the value of persistent state. History folding reaches 22.5\% Pass@1 but only 32.5\% Pass@3, confirming that compression alone cannot preserve task-critical facts. Self-describing step outputs reach 25.0\% Pass@1, indicating that grounded observations and action intents help context reuse. Full \conact reaches 40.0\% Pass@1 / 62.5\% Pass@3, far exceeding any single-component variant. 

\insight{The three components are complementary: folding controls context growth, memory preserves exact facts, and self-description grounds both.}

\subsection{Case Study and Error Analysis}
\label{sec:case-error}
\vspace{-0.25em}

\definecolor{casebenchpurplebg}{HTML}{ECE5F0}
\definecolor{casebenchbluebg}{HTML}{EEF2F3}
\definecolor{casefoldpurple}{HTML}{A983BD}
\definecolor{casememoryorange}{HTML}{F0A65C}
\newcommand{\benchbg}[2]{\begingroup\setlength{\fboxsep}{0.5pt}\colorbox{#1}{\textbf{#2}}\endgroup}
\newcommand{\casefoldhl}[1]{\begingroup\setlength{\fboxsep}{0.5pt}\colorbox{casefoldpurple}{\textcolor{white}{\normalfont\itshape #1}}\endgroup}
\newcommand{\casememhl}[1]{\begingroup\setlength{\fboxsep}{0.5pt}\colorbox{casememoryorange}{\textcolor{white}{\normalfont\itshape #1}}\endgroup}

\begin{figure*}[t]
    \centering
    \includegraphics[width=0.94\textwidth]{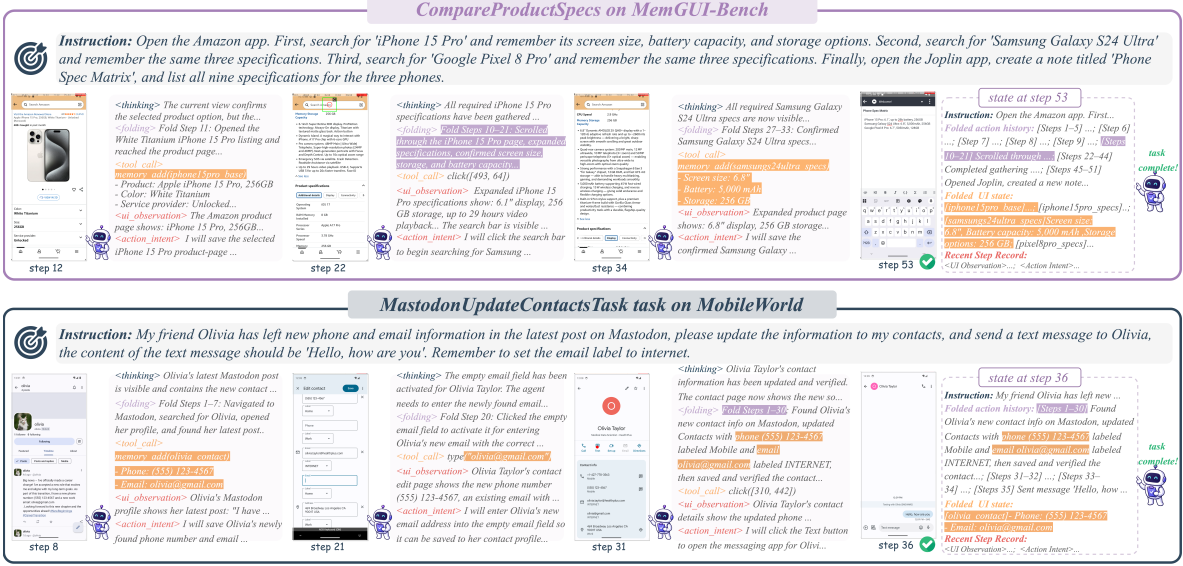}
    \caption[Successful zero-shot trajectories of \ourmethod-235B on MemGUI-Bench and MobileWorld.]{Successful zero-shot trajectories of \ourmethod-235B on \benchbg{casebenchpurplebg}{MemGUI-Bench} and \benchbg{casebenchbluebg}{MobileWorld}. Read rows left to right: screenshots are milestones, snippets are \conact outputs, \casefoldhl{folded action history}/\casememhl{memory writes} mark compressed steps/saved facts, and dashed boxes show compact state.}
    \label{fig:case-study}
\end{figure*}

\paragraph{Case study.}
Figure~\ref{fig:case-study} shows two successful zero-shot \ourmethod-235B trajectories. On MemGUI-Bench, the agent compares Amazon product specifications and records facts in Joplin; on MobileWorld, it carries contact information across Mastodon, Contacts, and Messages. Both cases show cross-app \conact behavior: writing critical facts before app transitions, folding completed subtasks, and preserving recent details. Full base-vs.-MemGUI trajectory comparisons are in Appendix~\ref{sec:appendix-trajectory-comparison}.

\begin{figure}[h]
  \centering
  \includegraphics[width=0.76\linewidth]{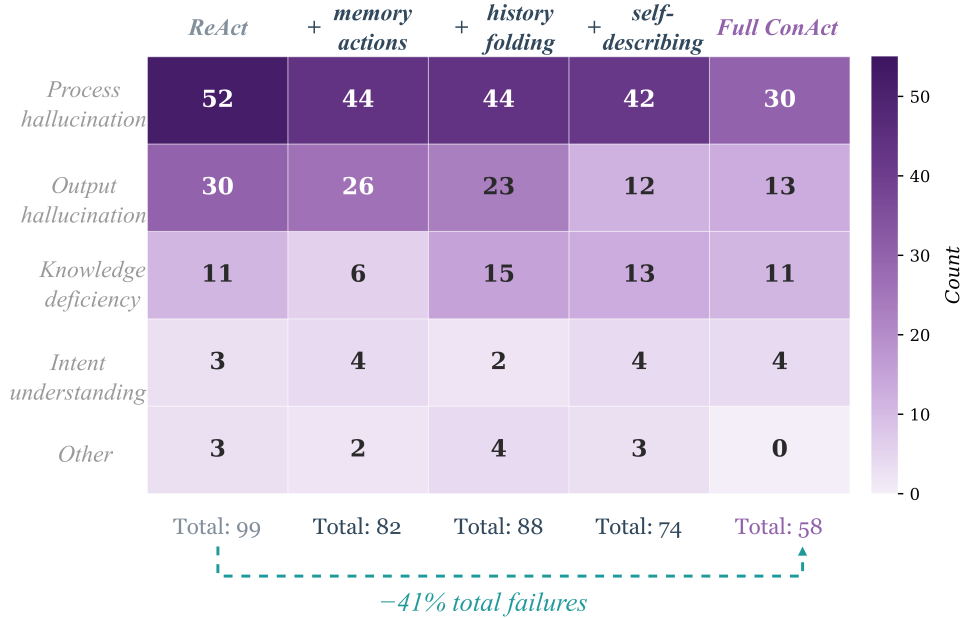}
  \caption[Failure category counts across ablation variants.]{Failure category counts for zero-shot \llmname{Qwen3-VL-235B-Thinking} across five ablation variants on MemGUI-Bench-40 (3 attempts $\times$ 40 tasks). Categories follow the MemGUI-Bench failure taxonomy~\cite{liu2026memgui}. Full \conact reduces total failures by 41\% (99$\to$58), mainly in process and output hallucination.}
  \label{fig:failure-heatmap}
\end{figure}

\paragraph{Error analysis.}
Following the MemGUI-Bench taxonomy~\cite{liu2026memgui}, Figure~\ref{fig:failure-heatmap} groups failures into process hallucination, output hallucination, knowledge deficiency, intent misunderstanding, and other errors. Full \conact reduces total failures by 41\% (99$\to$58), mainly through process hallucination ($-$22, $-$42\%) and output hallucination ($-$17, $-$57\%). Knowledge deficiency, intent misunderstanding, and other errors remain roughly stable, suggesting that \conact mainly prevents progress loss and fabricated UI facts, while remaining errors require stronger knowledge, intent understanding, and environment robustness. Taxonomy details and representative examples are in Appendix~\ref{sec:appendix-error-analysis}.

\insight{\conact mainly reduces context-induced hallucinations; the remaining ceiling lies in knowledge, intent understanding, and environment robustness.}

\section{Conclusion}
\label{sec:conclusion}

We presented \ourmethod, an end-to-end long-horizon mobile GUI agent that manages context inside the action policy through \conact, unifying history folding, UI memory actions, and self-describing step outputs. Zero-shot \conact sets a new state of the art on MemGUI-Bench with \llmname{Qwen3-VL-235B-Thinking}. We further construct MemGUI-3K, enabling MemGUI-8B-SFT to achieve the best open-data 8B performance and generalize to MobileWorld.

\section*{Limitations}

Our experiments focus on Android-style mobile GUI environments. Extending \conact to iOS, desktop, and web interfaces remains future work.

\bibliographystyle{plainnat}
\bibliography{custom}

\clearpage
\appendix

\noindent\textbf{Appendix organization.}
We place background discussion first, then benchmark and dataset details, followed by training, offline evaluation, qualitative examples, and failure analysis protocols; complete prompt templates are collected at the end.

\section{Related Work}
\label{sec:related-work}
\vspace{-0.25em}

\paragraph{Mobile GUI agents.}
Mobile GUI agents built on multimodal models generally follow two paradigms. 
\emph{Agentic frameworks} combine a backbone MLLM with planning, memory, and grounding modules~\cite{rawles2025androidworld,wang2024mobile,wang2025mobileagentselfevolving,agashe2025agent}, often around proprietary backbones such as \llmname{Gemini-2.5-Pro}~\cite{comanici2025gemini}; they can perform strongly but are complex and difficult to reproduce. 
\emph{End-to-end models}~\cite{gao2026ui,zhou2025mai,hong2024cogagent,xu2026mobile} directly map screenshots to actions, offering simplicity and deployability, but typically rely on passive context mechanisms: Action-Thought traces~\cite{zhang2025appagent,gu2025ui-venus,xu2026mobile,bai2025qwen3}, multi-turn prompts~\cite{qin2025ui}, rule-based aggregation~\cite{zheng2024seeact}, or no-history prompting~\cite{hong2024cogagent}. 
These mechanisms either let prompts grow with task horizon or lose UI-derived facts needed across screens and apps~\cite{liu2026memgui,kong2025mobileworld}. 
\ourmethod brings proactive context management into the end-to-end paradigm itself.

\paragraph{Context management in LLM agents.}
Long-context LLM agents manage histories through external memory or prompt-internal curation. 
External memory maintains a separate store, such as retrieval-augmented databases~\cite{lewis2020retrieval,chhikara2025mem0,xu2026mem}, hierarchical memory~\cite{zhang2026g}, or experiential replay~\cite{zhao2024expel}. 
Prompt-internal curation edits the in-task context through truncation, summarization, or model-controlled compression, as in recent web-agent methods such as MEM1~\cite{zhou2025mem1}, MemAgent~\cite{yu2025memagent}, and AgentFold~\cite{ye2025agentfold}. 
Unlike text-only web settings, mobile GUI control must preserve UI-derived facts, such as prices, codes, and copied text, verbatim across screen and app transitions. 
\conact addresses this by treating context curation as joint action over structured history, UI state, and recent-step fields within the same end-to-end policy.

\section{Benchmark and Online Evaluation Protocol}
\label{sec:appendix-eval-protocol}

\paragraph{Benchmark details.}
MemGUI-Bench~\cite{liu2026memgui} contains 128 tasks across 26 Android apps, of which 89.8\% are explicitly memory-intensive. Tasks are partitioned by golden-step count into Easy (1--20, 48 tasks), Medium (21--40, 42 tasks), and Hard (41+, 38 tasks); 78.1\% are cross-app, with an average length of 36.2 steps. The 40-task subset MemGUI-Bench-40 is reserved for ablations. MobileWorld~\cite{kong2025mobileworld} GUI-Only contains 117 tasks in a containerized environment with disjoint apps (Mattermost, Mastodon, Mall4Uni), averaging 27.8 steps per task with 62.2\% cross-app.

\paragraph{Compared agents.}
On MemGUI-Bench, we compare against agentic frameworks powered by \llmname{Gemini-2.5-Pro}: Agent-S2~\cite{agashe2025agent}, M3A/T3A~\cite{rawles2025androidworld}, Mobile-Agent-E~\cite{wang2025mobileagentselfevolving}, Mobile-Agent-V2~\cite{wang2024mobile}, SeeAct~\cite{zheng2024seeact}, and AppAgent~\cite{zhang2025appagent}; and end-to-end models including \llmname{Qwen3-VL} 8B/32B/235B variants~\cite{bai2025qwen3}, \llmname{GUI-Owl-1.5} 8B/32B~\cite{xu2026mobile}, \llmname{UI-TARS-1.5-7B}~\cite{qin2025ui}, and \llmname{UI-Venus-7B}~\cite{gu2025ui-venus}. On MobileWorld, we additionally include \llmname{ScaleCUA-7B}~\cite{ScaleCUA2025}, \llmname{OpenMobile} 7B/8B~\cite{OpenMobile2025}, and \llmname{ClawGUI-2B}~\cite{tang2026clawgui}.

\paragraph{MemGUI-Bench evaluation.}
For MemGUI-Bench and MemGUI-Bench-40, each task is evaluated with up to 3 independent attempts. For a task with golden-step count $g$, the maximum interaction budget is
\[
B_{\mathrm{MemGUI}}(g) = 2.5g + 1,
\]
rounded to an integer in implementation. This adaptive budget gives longer tasks proportionally more room while preventing unlimited exploration. We report Pass@$k$ for $k{\in}\{1,3\}$, where a task is counted as successful if any of the first $k$ attempts is judged successful by MemGUI-Eval.

\paragraph{MobileWorld evaluation.}
For MobileWorld GUI-Only, we follow the benchmark protocol and run a single attempt per task with a 50-step limit. We report Pass@1 success rate (SR), where task success is determined by the MobileWorld backend verifier.

\paragraph{Memory-oriented metrics.}
Besides success rate, MemGUI-Bench reports memory-oriented metrics. IRR (Information Retention Rate) measures whether task-critical information units are retained and correctly used. MTPR (Memory-Task Proficiency Ratio) measures success on explicitly memory-intensive tasks. These metrics are computed by the MemGUI-Eval pipeline~\cite{liu2026memgui}.

\section{MemGUI-3K Dataset Details}
\label{sec:appendix-dataset}

This appendix provides additional details for the MemGUI-3K construction pipeline summarized in \S\ref{sec:dataset}.

\subsection{Task Expansion}
\label{sec:appendix-task-expansion}

We start from the 128 seed tasks of MemGUI-Bench~\cite{liu2026memgui}, which are long-horizon mobile GUI tasks with strong cross-step dependencies. We expand them through three complementary strategies.

\begin{itemize}
    \item \textbf{Entity substitution} preserves the original task pipeline but replaces task-specific entities, such as product names, search queries, contacts, notes, locations, or app-specific targets. This yields 456 expanded tasks.
    \item \textbf{Memory-operation augmentation} explicitly constructs tasks requiring not only \memaction{memory\_add}, but also \memaction{memory\_update} and \memaction{memory\_delete}. This yields 289 memory-operation-focused tasks.
    \item \textbf{Task simplification} decomposes complex expanded tasks into shorter, single-objective variants while preserving the need for structured context management. This yields 6{,}558 simplified tasks.
\end{itemize}

The three strategies produce a combined pool of 7{,}303 tasks, of which 5{,}293 are selected for rollout. The final dataset contains three source types: simplified tasks, entity-substitution tasks, and memory-operation augmentation tasks.

\begin{table}[h]
\centering
\footnotesize
\begin{adjustbox}{max width=0.70\linewidth}
\begin{tabular}{lrrr}
\toprule
Source prefix & Total & Train & Test \\
\midrule
S: simplified & 2{,}566 (86.8\%) & 2{,}309 & 257 \\
E: entity substitution & 234 (7.9\%) & 208 & 26 \\
M: memory operations & 156 (5.3\%) & 144 & 12 \\
\bottomrule
\end{tabular}
\end{adjustbox}
\caption{Task-source composition of MemGUI-3K.}
\label{tab:appendix-prefix}
\end{table}

\subsection{Teacher Rollout and Trajectory Filtering}
\label{sec:appendix-rollout-filtering}

All rollouts are collected with \llmname{Qwen3-VL-235B-Thinking} using the full 5-part \conact protocol described in \S\ref{sec:method}. Each generated task is executed once in the snapshot-based Android environment of MemGUI-Bench. The environment is reset before each task and supports UI actions, such as \texttt{click}, \texttt{swipe}, \texttt{type}, \texttt{wait}, and \texttt{system\_button}, as well as memory actions, i.e., \memaction{memory\_add}, \memaction{memory\_update}, and \memaction{memory\_delete}. For a generated task with golden-step count $g$, the teacher rollout budget is also set to $2.5g+1$ steps, following the MemGUI-Bench evaluation protocol. Each rollout stops when the agent emits \texttt{terminate} or reaches this task-specific budget.

We first retain only successful trajectories according to the final decision of MemGUI-Eval~\cite{liu2026memgui}. MemGUI-Eval follows a progressive scrutiny pipeline: it first screens task completion from the goal, action logs, and final screenshots; then performs semantic analysis with generated step descriptions and memory-related checks; and finally verifies critical visual information against requested screenshots. Among 5{,}293 rollout trajectories, 2{,}959 receive a positive final decision. We then remove one abnormal 321-step outlier from an early collection run and two successful trajectories involving low-frequency apps with at most 8 occurrences. This produces the final 2{,}956 MemGUI-3K trajectories.

\begin{table}[h]
\centering
\footnotesize
\begin{adjustbox}{max width=0.62\linewidth}
\begin{tabular}{lrr}
\toprule
Filtering stage & Removed & Remaining \\
\midrule
Successful trajectories & -- & 2{,}959 \\
Remove abnormal 321-step outlier & 1 & 2{,}958 \\
Remove low-frequency-app tasks & 2 & 2{,}956 \\
\bottomrule
\end{tabular}
\end{adjustbox}
\caption{Trajectory-level filtering for MemGUI-3K.}
\label{tab:appendix-filtering}
\end{table}

The final trajectories are split 90\%/10\% at the trajectory level with seed 42. We verify zero overlap against the 128-task MemGUI-Bench evaluation set by checking both task identifiers and task descriptions. Since MemGUI-Bench-40 is a subset of MemGUI-Bench, it is also disjoint from MemGUI-3K.

\begin{table}[h]
\centering
\footnotesize
\begin{adjustbox}{max width=0.68\linewidth}
\begin{tabular}{lrrr}
\toprule
Split & Trajectories & SFT samples & Avg. reasonable steps \\
\midrule
Train & 2{,}661 & 57{,}951 & 21.8 \\
Test & 295 & 6{,}479 & 22.0 \\
\midrule
Total & 2{,}956 & 64{,}430 & 21.8 \\
\bottomrule
\end{tabular}
\end{adjustbox}
\caption{Train/test split and reasonable-step SFT samples.}
\label{tab:appendix-split}
\end{table}

\subsection{Step-Level Reasonableness Filtering}
\label{sec:appendix-step-filtering}

Successful trajectories may still contain redundant, exploratory, or counterproductive steps that are later corrected. To avoid training the student on such actions, we extend MemGUI-Eval with step-level reasonableness analysis. The evaluator labels each step as \emph{reasonable} or \emph{unreasonable}, attaches a natural-language explanation, and assigns an impact label (\emph{positive}, \emph{negative}, or \emph{neutral}). Only steps labeled as reasonable are converted into SFT samples.

The following instruction block is appended to the MemGUI-Eval final-decision prompt. The evaluator receives the task description, action logs, generated step descriptions, and screenshots, and returns both the trajectory-level judgment and per-step reasonableness annotations in a single JSON response.

\begin{Verbatim}[fontsize=\footnotesize,breaklines=true,breakanywhere=true]
STEP REASONABLENESS ANALYSIS

In addition to the overall task success/failure judgment, provide a step-level reasonableness analysis for every step in the trajectory.

Return:
- reasonable_steps: an array of step numbers that are reasonable and contribute to task completion.
- unreasonable_steps: an array of step numbers that are incorrect, redundant, or counterproductive.
- step_analysis: an object keyed by step number. For each step, include:
    - reasonableness: "reasonable" or "unreasonable"
    - explanation: why the step is reasonable or unreasonable
    - impact: "positive", "negative", or "neutral"

Notes:
- Successful trajectories may still contain unreasonable steps that were later corrected.
- Failed trajectories may still contain reasonable steps that move toward the goal.
\end{Verbatim}

\subsection{Training Data Example}
\label{sec:appendix-training-data-example}

\begin{figure*}[p]
  \centering
  \includegraphics[height=0.94\textheight,width=0.92\textwidth,keepaspectratio]{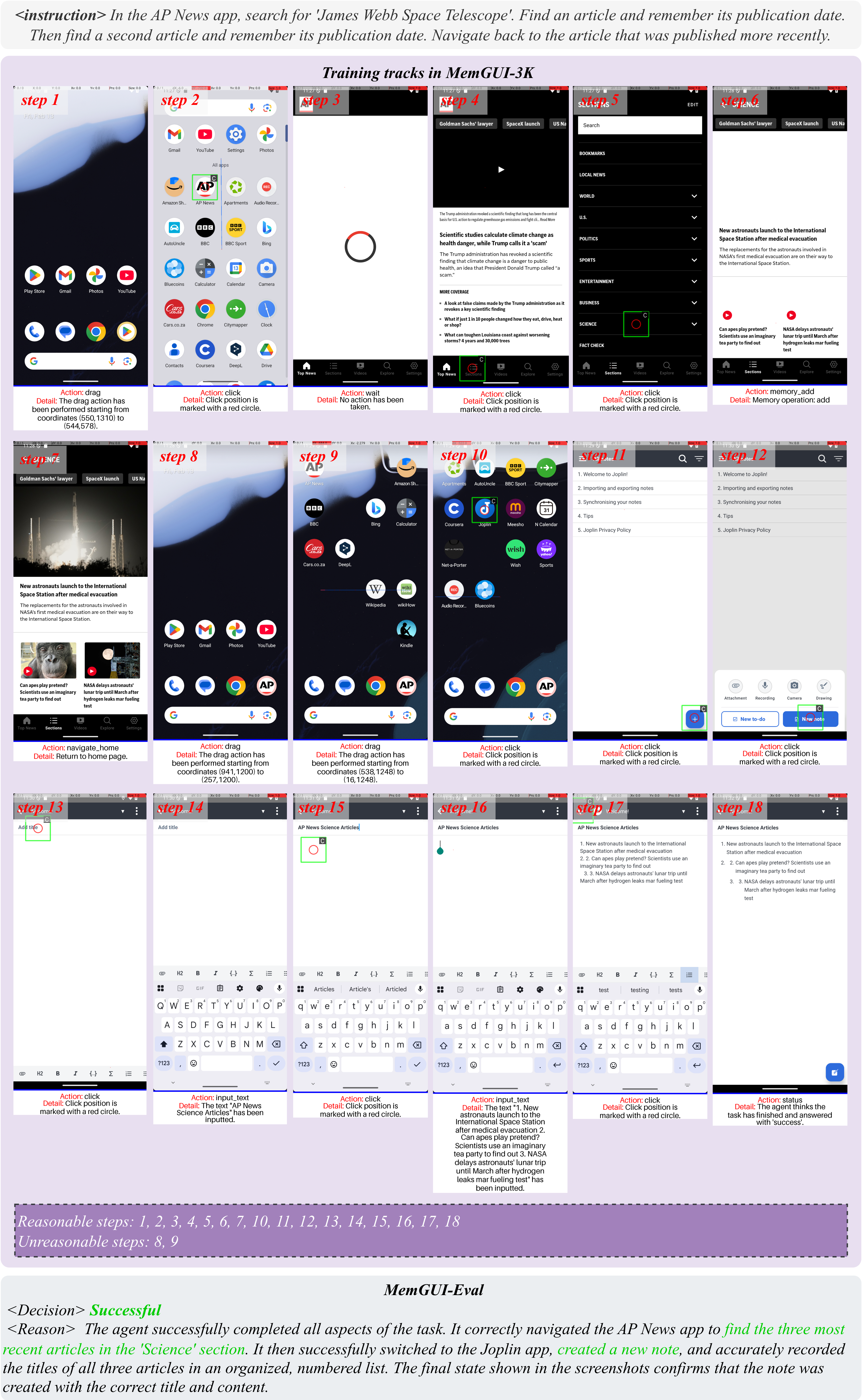}
  \caption{Training-data example from MemGUI-3K. The trajectory contains step-level reasonableness labels, explanations, and impact annotations. Reasonable steps provide supervised \conact targets; unreasonable steps illustrate redundant, incorrect, or counterproductive behavior filtered out of SFT samples.}
  \label{fig:appendix-training-data-example}
\end{figure*}

Figure~\ref{fig:appendix-training-data-example} shows one MemGUI-3K trajectory with step-level reasonableness annotations. Reasonable steps are converted into SFT targets, while unreasonable steps are retained in the trajectory-level record for analysis but excluded from the step-level training split.

\clearpage

\subsection{Dataset Statistics}
\label{sec:appendix-dataset-statistics}

MemGUI-3K contains long-horizon trajectories with an average of 28.8 steps and a median of 25 steps. After step-level filtering, the average number of reasonable steps is 21.8, corresponding to a 75.7\% reasonable-step ratio.

\begin{table}[h]
\centering
\footnotesize
\begin{adjustbox}{max width=0.48\linewidth}
\begin{tabular}{lrr}
\toprule
Metric & Mean & Median \\
\midrule
All steps & 28.8 & 25 \\
Reasonable steps & 21.8 & 20 \\
\bottomrule
\end{tabular}
\end{adjustbox}
\caption{Trajectory length statistics under all-step and reasonable-step views.}
\label{tab:appendix-length}
\end{table}

MemGUI-3K contains both step-level and span-level folding annotations. Step-level folds distill the latest step, while span-level folds compress a completed multi-step subtask into a reusable record.

\begin{table}[h]
\centering
\footnotesize
\begin{adjustbox}{max width=0.56\linewidth}
\begin{tabular}{lr}
\toprule
Metric & Value \\
\midrule
Step-level folds & 56{,}447 \\
Span-level folds & 17{,}664 \\
Span-level fold ratio & 23.8\% \\
Average span length & 6.25 steps \\
Median span length & 4 steps \\
95th percentile span length & 18 steps \\
Maximum span length & 118 steps \\
Trajectories with $\geq$1 span-level fold & 88.7\% \\
\bottomrule
\end{tabular}
\end{adjustbox}
\caption{Folding-granularity statistics in MemGUI-3K.}
\label{tab:appendix-folding}
\end{table}

Context actions appear frequently in MemGUI-3K. On average, each trajectory contains 1.17 \memaction{memory\_add} actions, 0.03 \memaction{memory\_update} actions, and 0.04 \memaction{memory\_delete} actions. At the trajectory level, 65.1\% of trajectories invoke at least one memory action.

MemGUI-3K spans 26 Android apps across 7 functional categories. The most frequent apps include Joplin, Amazon, Bing, Calculator, Apartments.com, Messages, Citymapper, AutoUncle, Calendar, and AP News. We normalize app-name variants before counting, e.g., \texttt{joplin}/\texttt{Joplin}, \texttt{bing}/\texttt{Bing}, and \texttt{messages}/\texttt{Messages}.

\subsection{Dataset Format}
\label{sec:appendix-dataset-format}

MemGUI-3K is released in two complementary formats. The trajectory-level format preserves complete rollouts, evaluation metadata, IRR statistics, token usage, screenshots, and step-level reasonableness annotations, supporting offline analysis of memory operations, folding granularity, token cost, and recovery patterns. The step-level format converts all reasonable steps into ms-swift-compatible SFT samples. Each sample contains the shared system prompt, the user message with structured context state and current screenshot, and the gold 5-part \conact response. Only steps labeled as reasonable are included in the SFT split.

The SFT data follows the standard chat-style multimodal format. A schematic example is shown below:
\begin{Verbatim}[fontsize=\scriptsize,breaklines,breakanywhere]
{
  "messages": [
    {"role": "system", "content": "..."},
    {"role": "user", "content":
      "<image>### User Query\n...### Folded Action History\n...### Folded UI State\n...### Recent Step Record\n..."},
    {"role": "assistant",
     "content": "<thinking>...</thinking>\n<folding>...</folding>\n<tool_call>...</tool_call>\n<ui_observation>...</ui_observation>\n<action_intent>...</action_intent>"}
  ],
  "images": ["data:image/png;base64,..."]
}
\end{Verbatim}

\section{Training Setup and Dynamics}
\label{sec:appendix-train}

\subsection{Training Setup}
\label{sec:appendix-training-setup}

MemGUI-8B-SFT is initialized from \llmname{Qwen3-VL-8B-Instruct} and fine-tuned on the 57{,}951 reasonable-step training samples of MemGUI-3K using LoRA SFT with ms-swift~\cite{zhao2024swift}. Each sample contains the shared system prompt, the user message consisting of the structured context state and current screenshot, and the gold 5-part \conact response. Training is conducted on eight 80GB GPUs and takes 535 minutes.

We use bfloat16 precision and train for one epoch with AdamW fused optimizer, cosine learning-rate scheduling, and a warmup ratio of 0.05. The per-device training batch size is 2 with gradient accumulation of 8, giving an effective batch size of 128 across 8 GPUs. We set the learning rate to $1\times10^{-4}$, weight decay to 0.1, maximum gradient norm to 1.0, and maximum sequence length to 32{,}768. LoRA uses rank 8, alpha 32, dropout 0.05, and is applied to all linear language-model projection modules. We use seed 42, four dataloader workers, and report training logs to Weights \& Biases.

\begin{table}[h]
\centering
\footnotesize
\begin{adjustbox}{max width=0.72\linewidth}
\begin{tabular}{ll}
\toprule
Hyperparameter & Value \\
\midrule
Base model & \llmname{Qwen3-VL-8B-Instruct} \\
Trainable parameters & LoRA adapters \\
Training samples & 57{,}951 reasonable steps \\
Epochs & 1 \\
Precision & bfloat16 \\
Optimizer & AdamW fused \\
LR scheduler & cosine \\
Learning rate & $1\times10^{-4}$ \\
Warmup ratio & 0.05 \\
Weight decay & 0.1 \\
Max grad norm & 1.0 \\
Max sequence length & 32{,}768 \\
Per-device train batch size & 2 \\
Gradient accumulation steps & 8 \\
Effective batch size & 128 \\
LoRA rank / alpha / dropout & 8 / 32 / 0.05 \\
LoRA target modules & all linear LM projections \\
Dataloader workers & 4 \\
Seed & 42 \\
Hardware & 8$\times$80GB \\
Training time & 535 minutes \\
Software & ms-swift, Transformers 4.57.0, Python 3.10 \\
\bottomrule
\end{tabular}
\end{adjustbox}
\caption{Training hyperparameters for MemGUI-8B-SFT.}
\label{tab:appendix-training-hparams}
\end{table}

\subsection{Training Dynamics}
\label{sec:appendix-training-dynamics}

\begin{figure}[h]
  \centering
  \includegraphics[width=0.84\linewidth]{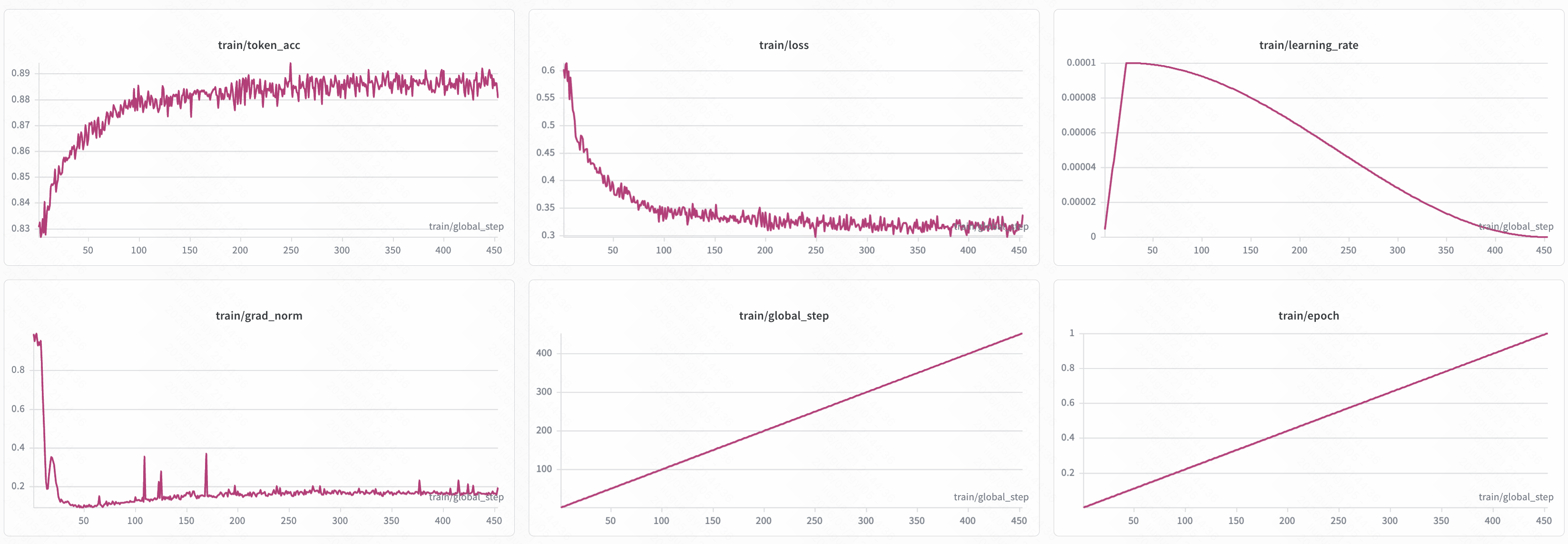}
  \caption{Training dynamics of MemGUI-8B-SFT. We report token accuracy, training loss, learning rate, gradient norm, global step, and epoch during LoRA SFT.}
  \label{fig:appendix-training-curves}
\end{figure}

Figure~\ref{fig:appendix-training-curves} shows the training dynamics of MemGUI-8B-SFT. The token accuracy steadily increases and the training loss decreases during the single-epoch LoRA SFT run, while the learning rate follows the configured warmup and cosine decay schedule.

\section{Offline Skill Evaluation Protocol}
\label{sec:appendix-offline-skill}

This section details the offline skill evaluation protocol used in Table~\ref{tab:offline-eval}. The implementation is provided in \texttt{memgui3k\_test\_eval/} and has three stages: \texttt{run\_offline\_eval.py} performs gold-context step-level inference, \texttt{evaluate\_predictions.py} computes automatic metrics, and \texttt{compare\_results.py} builds the model-comparison summary. The reported run evaluates three models on the MemGUI-3K test split, which contains 295 trajectories and 6{,}479 reasonable-step samples.

\paragraph{Offline inference protocol.}
For each test sample, the inference script sends only the first two chat messages from \texttt{test\_sft.jsonl}, namely the system prompt and user message containing the task, structured context state, and screenshot. The gold assistant message is held out for evaluation. We use an OpenAI-compatible chat-completion API with temperature 0.01, maximum generation length 16{,}384 tokens, 32 concurrent workers, resumable JSONL output, and retry logic for transient API errors. The predictions are written as \texttt{\{task\_id, step, prediction, gold, usage\}} records, and the evaluator writes both an aggregate JSON report and per-step details.

\begin{table}[h]
\centering
\footnotesize
\renewcommand{\arraystretch}{1.04}
\setlength{\tabcolsep}{3.2pt}
\begin{adjustbox}{max width=0.78\linewidth}
\begin{tabular}{l ccc}
\toprule
& \multicolumn{3}{c}{\textbf{Model}} \\
\textbf{Metric} & \texttt{8B-Inst.} & \texttt{8B-SFT} & \texttt{235B-Th.} \\
\midrule
\multicolumn{4}{l}{\textit{\textbf{Overall Action Matching}}} \\
\quad Type Accuracy & 52.7 & \textbf{62.6} & 58.7 \\
\quad Match Accuracy & 28.2 & \textbf{35.6} & 33.1 \\
\midrule
\multicolumn{4}{l}{\textit{\textbf{UI Actions}}} \\
\quad Type Accuracy & 55.1 & \textbf{65.8} & 61.1 \\
\quad Match Accuracy & 29.2 & \textbf{36.3} & 33.9 \\
\quad\quad click & 18.7 & \textbf{26.7} & 24.2 \\
\quad\quad type & 53.1 & \textbf{69.5} & 67.3 \\
\quad\quad swipe & 41.2 & \textbf{45.4} & 44.5 \\
\quad\quad system\_button & 63.6 & \textbf{69.0} & 65.6 \\
\midrule
\multicolumn{4}{l}{\textit{\textbf{Memory Actions}}} \\
\quad Type Accuracy & 10.9 & \textbf{56.4} & 23.0 \\
\quad Match Accuracy & 10.6 & \textbf{51.5} & 22.4 \\
\quad Trigger F1 & 19.9 & \textbf{48.0} & 34.3 \\
\midrule
\multicolumn{4}{l}{\textit{\textbf{Folding}}} \\
\quad Deep Fold Ratio$^\dagger$ & 8.8 & \textbf{26.1} & 18.7 \\
\quad Range Acc (deep) & 45.2 & \textbf{58.9} & 56.7 \\
\quad Range Acc (shallow) & \textbf{98.2} & 86.2 & 92.3 \\
\midrule
\multicolumn{4}{l}{\textit{\textbf{Format Compliance}}} \\
\quad All Tags Correct & 94.9 & \textbf{99.9} & 99.2 \\
\bottomrule
\end{tabular}
\end{adjustbox}
\caption{Detailed aggregate metrics for gold-context step-level offline evaluation on the MemGUI-3K test set (295 trajectories, 6{,}479 steps). \textbf{8B-Inst.}~=~\llmname{Qwen3-VL-8B-Instruct} (zero-shot); \textbf{8B-SFT}~=~MemGUI-8B-SFT; \textbf{235B-Th.}~=~\llmname{Qwen3-VL-235B-Thinking} (zero-shot rerun). Match accuracy follows the implementation in Appendix~\ref{sec:appendix-offline-skill}: click/long-press within 14\% of the screen diagonal, type/answer exact match or token F1 $>$ 0.5, and memory add/update matching by description or content F1 $>$ 0.5. $^\dagger$Gold deep fold ratio is 23.2\%. \textbf{Bold}: best per row.}
\label{tab:offline-eval-detailed}
\end{table}

\paragraph{Action matching.}
The evaluator parses JSON from the \texttt{<tool\_call>} block and follows a ClawGUI-style two-level protocol~\cite{tang2026clawgui}. Type accuracy checks whether the predicted action type matches the gold action type. Match accuracy additionally checks action arguments. For \texttt{click} and \texttt{long\_press}, the predicted coordinate must fall within 14\% of the 1000$\times$1000 screen diagonal from the gold coordinate. For \texttt{swipe}, both start and end coordinates must fall within 1.5 times this threshold. For \texttt{type} and \texttt{answer}, the text must match exactly after stripping whitespace or achieve token-level F1 $>$ 0.5. For \texttt{system\_button} and \texttt{terminate}, the button/status argument must match exactly; \texttt{wait} succeeds whenever the action type matches.

\paragraph{Memory-action metrics.}
For memory actions, type accuracy evaluates whether the predicted operation is \texttt{memory\_add}, \texttt{memory\_update}, or \texttt{memory\_delete} when the gold output requires the same operation. For \texttt{memory\_delete}, match success requires exact \texttt{memory\_id} match. For \texttt{memory\_add} and \texttt{memory\_update}, match success requires either description token F1 $>$ 0.5 or content token F1 $>$ 0.5. The description criterion is used because numeric contents such as prices may have low token overlap even when the predicted memory intent is correct. Trigger precision, recall, and F1 treat any memory action as a binary event over all steps and evaluate whether the model invokes memory at the same steps as the gold response.

\paragraph{Folding metrics.}
The evaluator parses JSON from the \texttt{<folding>} block. Presence accuracy checks whether folding is present or absent consistently with gold. Range accuracy checks whether both predicted range endpoints are within $\pm2$ steps of the gold endpoints. A fold is considered span-level, or deep, if its range covers more than one step; otherwise it is shallow. We report predicted deep-fold ratio and separate range accuracy for shallow and deep folds, because a model can obtain a high overall range score by emitting mostly shallow folds.

\paragraph{Format compliance.}
Format compliance checks whether the output contains matched opening and closing tags for \texttt{<thinking>}, \texttt{<folding>}, \texttt{<tool\_call>}, \texttt{<ui\_observation>}, and \texttt{<action\_intent>}. The \texttt{<folding>} tag is treated as optional for step 1, matching the \conact protocol. The evaluator also computes a trajectory-level full-match rate, defined as the fraction of trajectories for which every evaluated step achieves action match success; we use this only as a diagnostic because the offline setting fixes gold context at each step rather than rolling out model-induced states.

\section{Trajectory Completion Comparisons}
\label{sec:appendix-trajectory-comparison}

\begin{figure*}[p]
  \centering
  \begin{subfigure}[t]{0.48\textwidth}
    \centering
    \includegraphics[width=\linewidth,height=0.64\textheight,keepaspectratio]{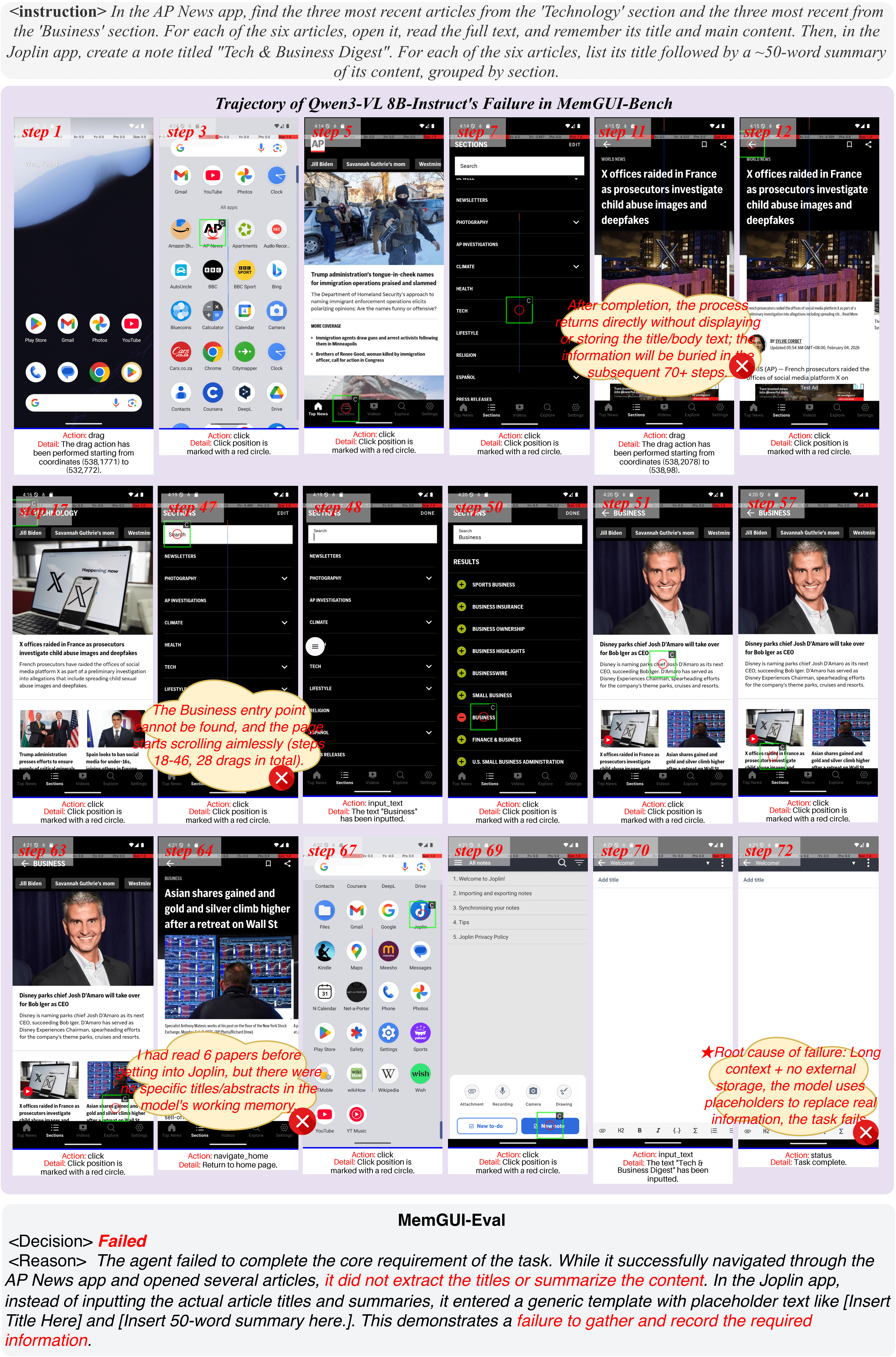}
    \caption{\llmname{Qwen3-VL-8B-Instruct} base.}
  \end{subfigure}
  \hfill
  \begin{subfigure}[t]{0.48\textwidth}
    \centering
    \includegraphics[width=\linewidth,height=0.64\textheight,keepaspectratio]{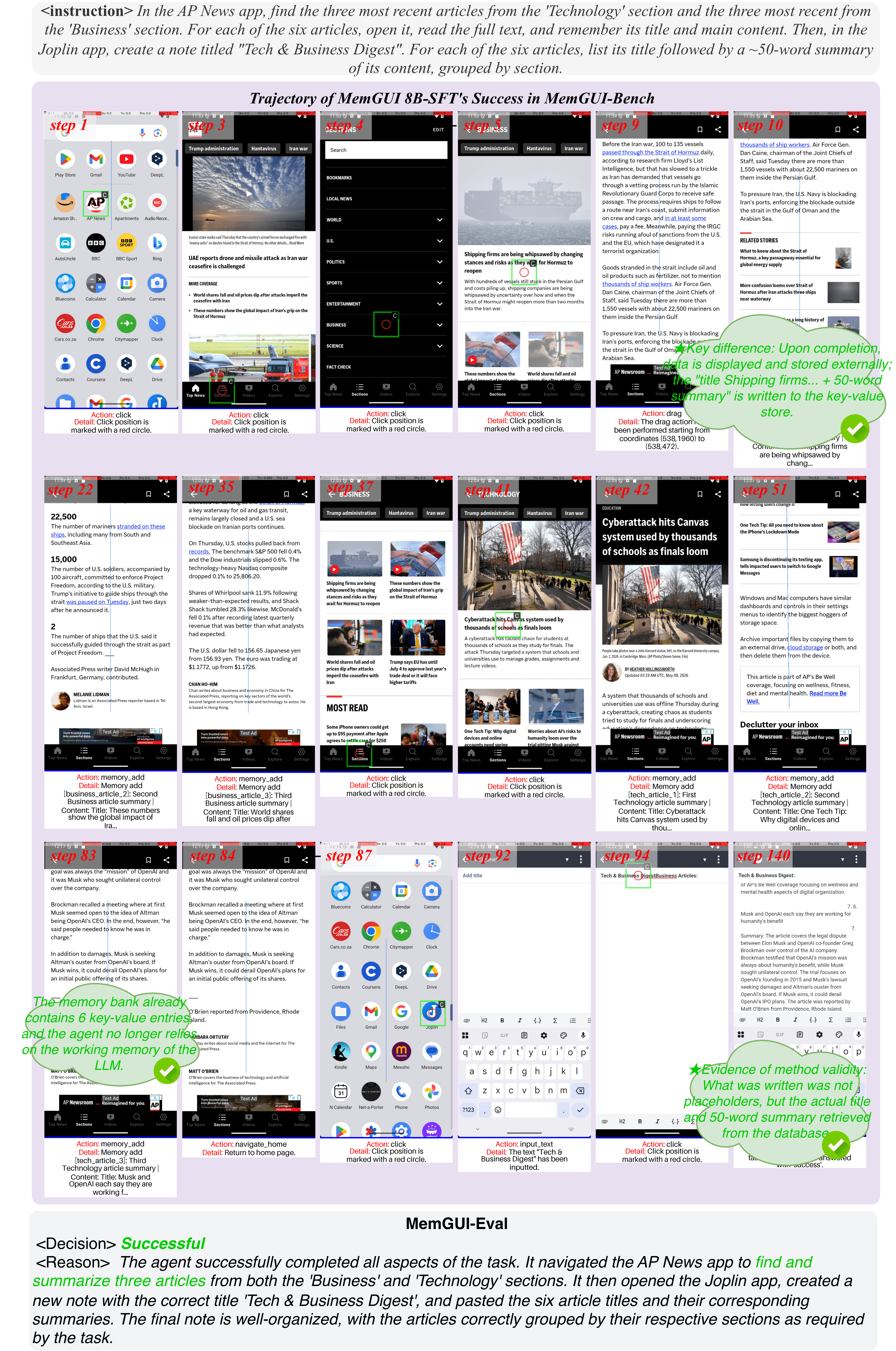}
    \caption{MemGUI-8B-SFT.}
  \end{subfigure}
  \caption[MemGUI-Bench trajectory comparison with an 8B backbone.]{MemGUI-Bench trajectory comparison on the same task with an 8B backbone. The base agent fails, while MemGUI-8B-SFT completes the task after learning \conact behavior from MemGUI-3K.}
  \label{fig:appendix-track-memguibench-8b}
\end{figure*}

\begin{figure*}[p]
  \centering
  \begin{subfigure}[t]{0.48\textwidth}
    \centering
    \includegraphics[width=\linewidth,height=0.64\textheight,keepaspectratio]{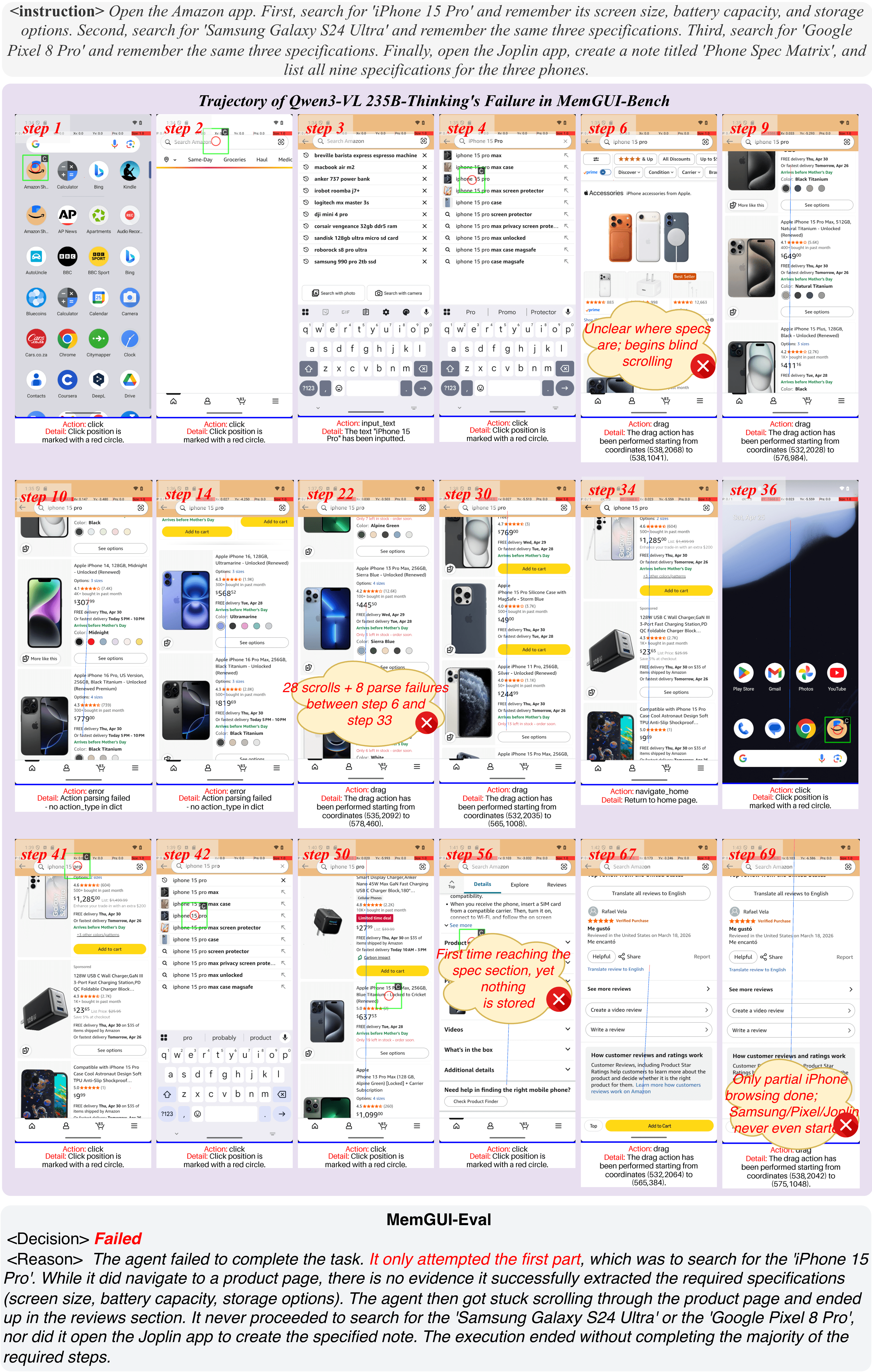}
    \caption{\llmname{Qwen3-VL-235B-Thinking} with ReAct-style prompting.}
  \end{subfigure}
  \hfill
  \begin{subfigure}[t]{0.48\textwidth}
    \centering
    \includegraphics[width=\linewidth,height=0.64\textheight,keepaspectratio]{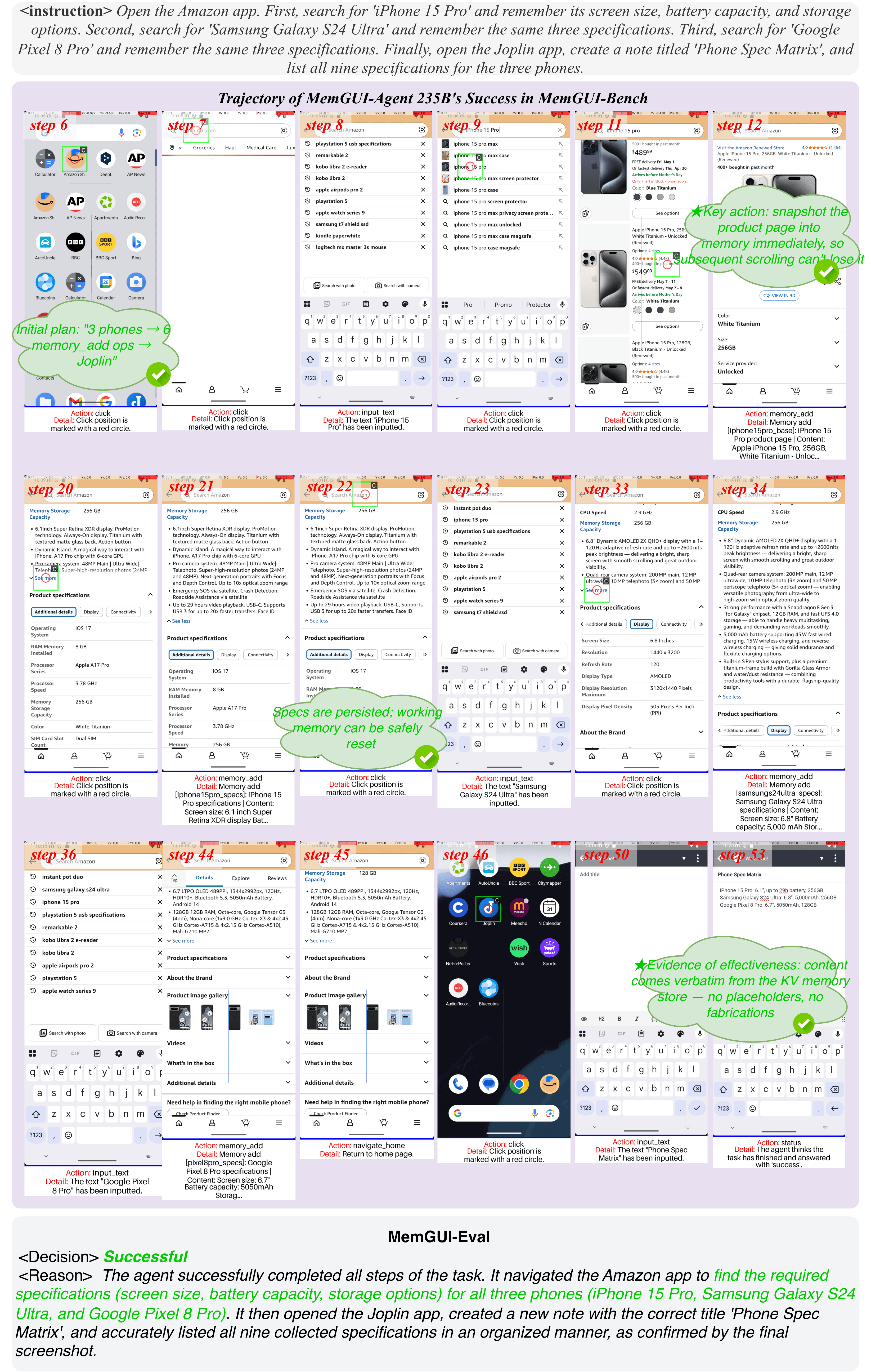}
    \caption{\ourmethod-235B with zero-shot \conact.}
  \end{subfigure}
  \caption[MemGUI-Bench trajectory comparison with a 235B backbone.]{MemGUI-Bench trajectory comparison on the same task with \llmname{Qwen3-VL-235B-Thinking}. \ourmethod-235B keeps the weights unchanged and changes only the prompting/action protocol.}
  \label{fig:appendix-track-memguibench-235b}
\end{figure*}

\begin{figure*}[p]
  \centering
  \begin{subfigure}[t]{0.48\textwidth}
    \centering
    \includegraphics[width=\linewidth,height=0.64\textheight,keepaspectratio]{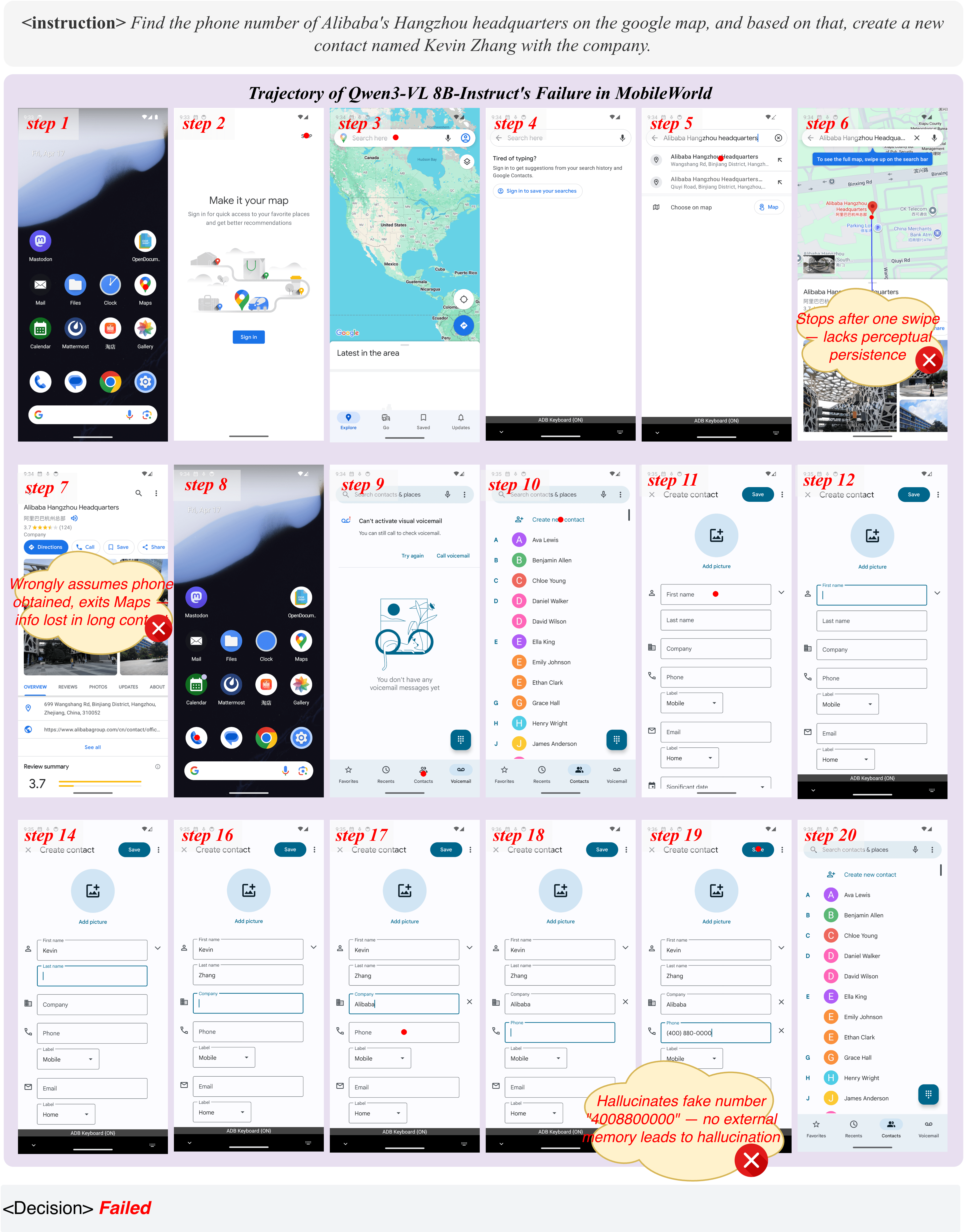}
    \caption{\llmname{Qwen3-VL-8B-Instruct} base.}
  \end{subfigure}
  \hfill
  \begin{subfigure}[t]{0.48\textwidth}
    \centering
    \includegraphics[width=\linewidth,height=0.64\textheight,keepaspectratio]{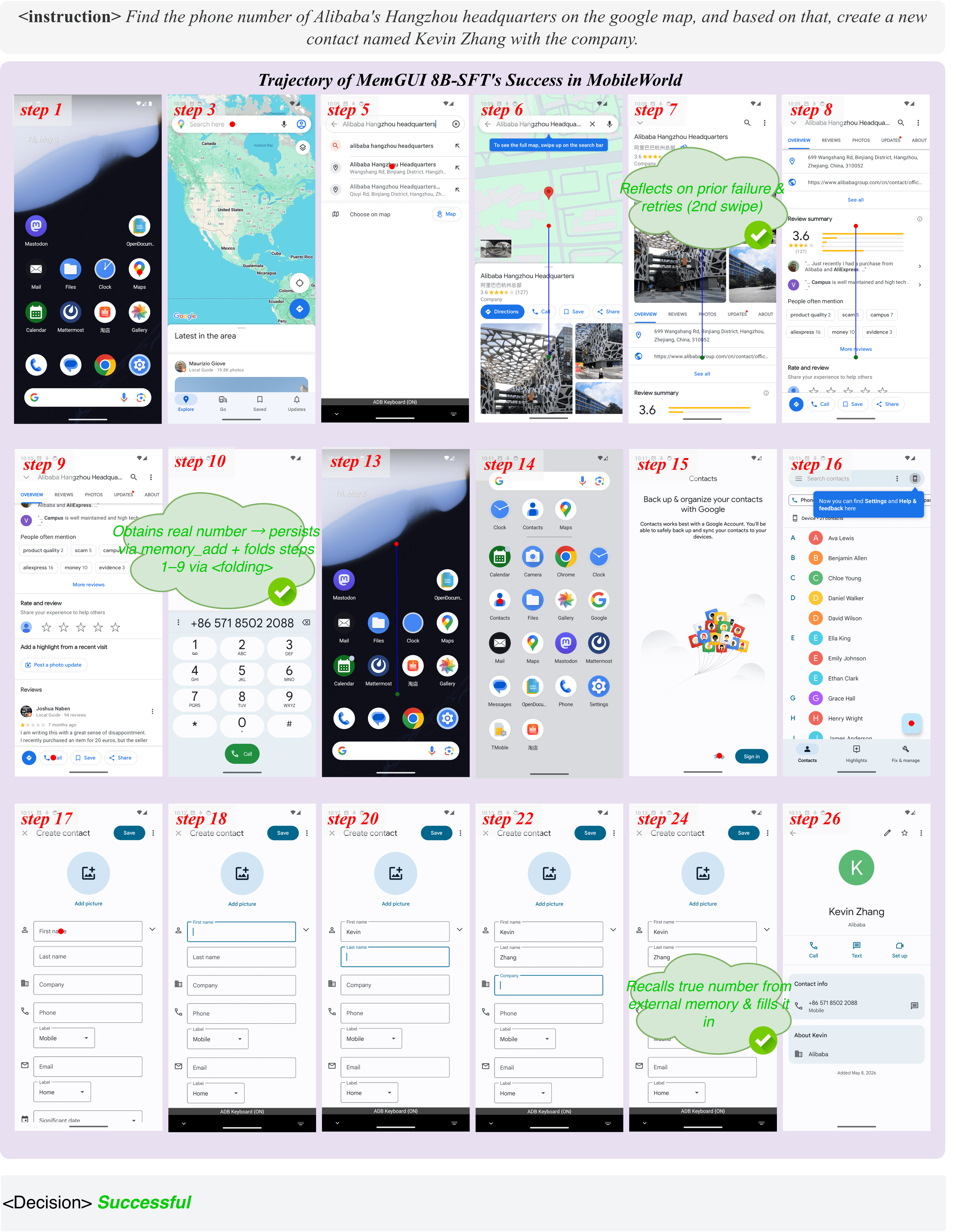}
    \caption{MemGUI-8B-SFT.}
  \end{subfigure}
  \caption[MobileWorld trajectory comparison with an 8B backbone.]{MobileWorld trajectory comparison on the same task with an 8B backbone. The example illustrates that learned context-management behavior transfers beyond the MemGUI-Bench source environment.}
  \label{fig:appendix-track-mobileworld-8b}
\end{figure*}

\begin{figure*}[p]
  \centering
  \begin{subfigure}[t]{0.48\textwidth}
    \centering
    \includegraphics[width=\linewidth,height=0.64\textheight,keepaspectratio]{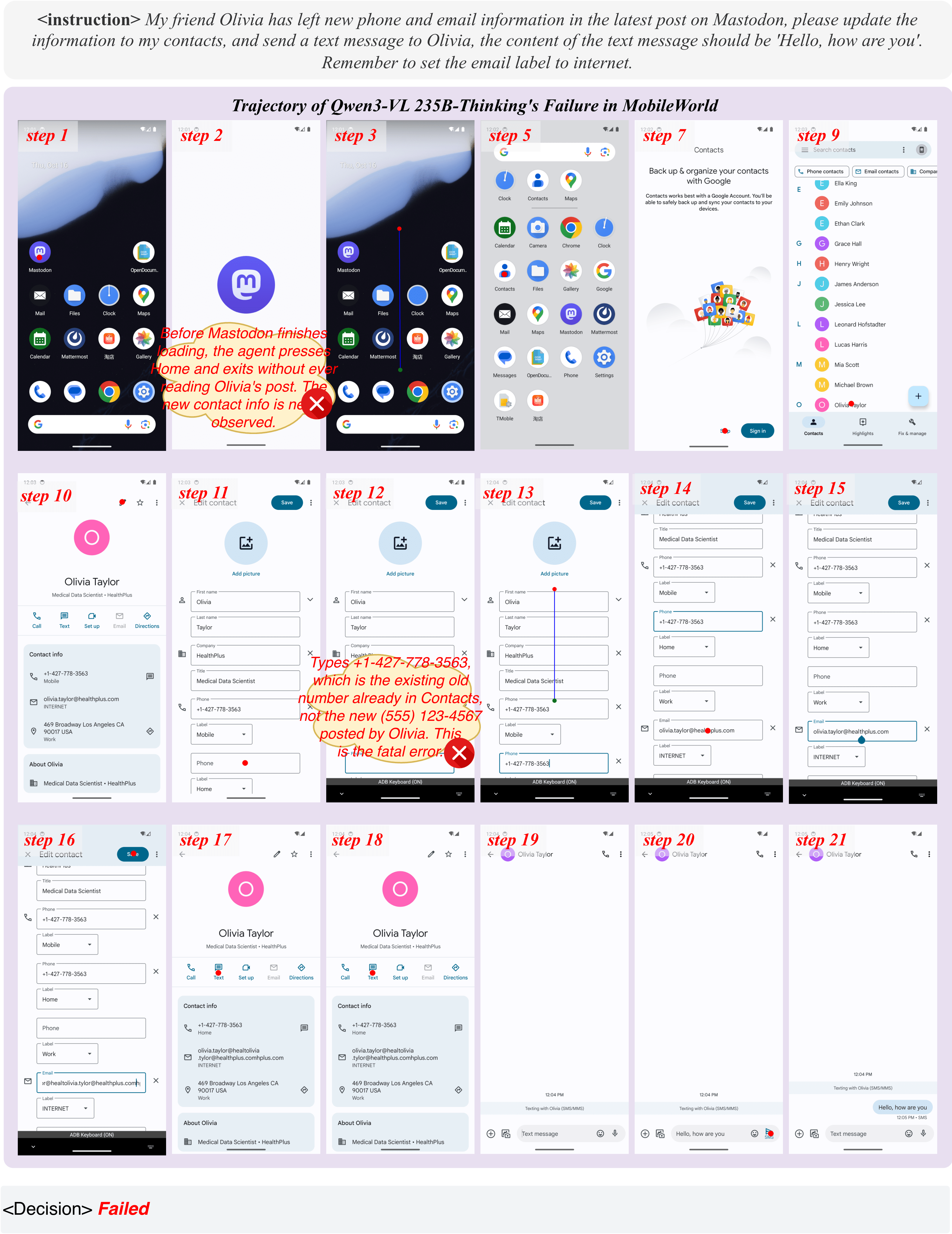}
    \caption{\llmname{Qwen3-VL-235B-Thinking} with ReAct-style prompting.}
  \end{subfigure}
  \hfill
  \begin{subfigure}[t]{0.48\textwidth}
    \centering
    \includegraphics[width=\linewidth,height=0.64\textheight,keepaspectratio]{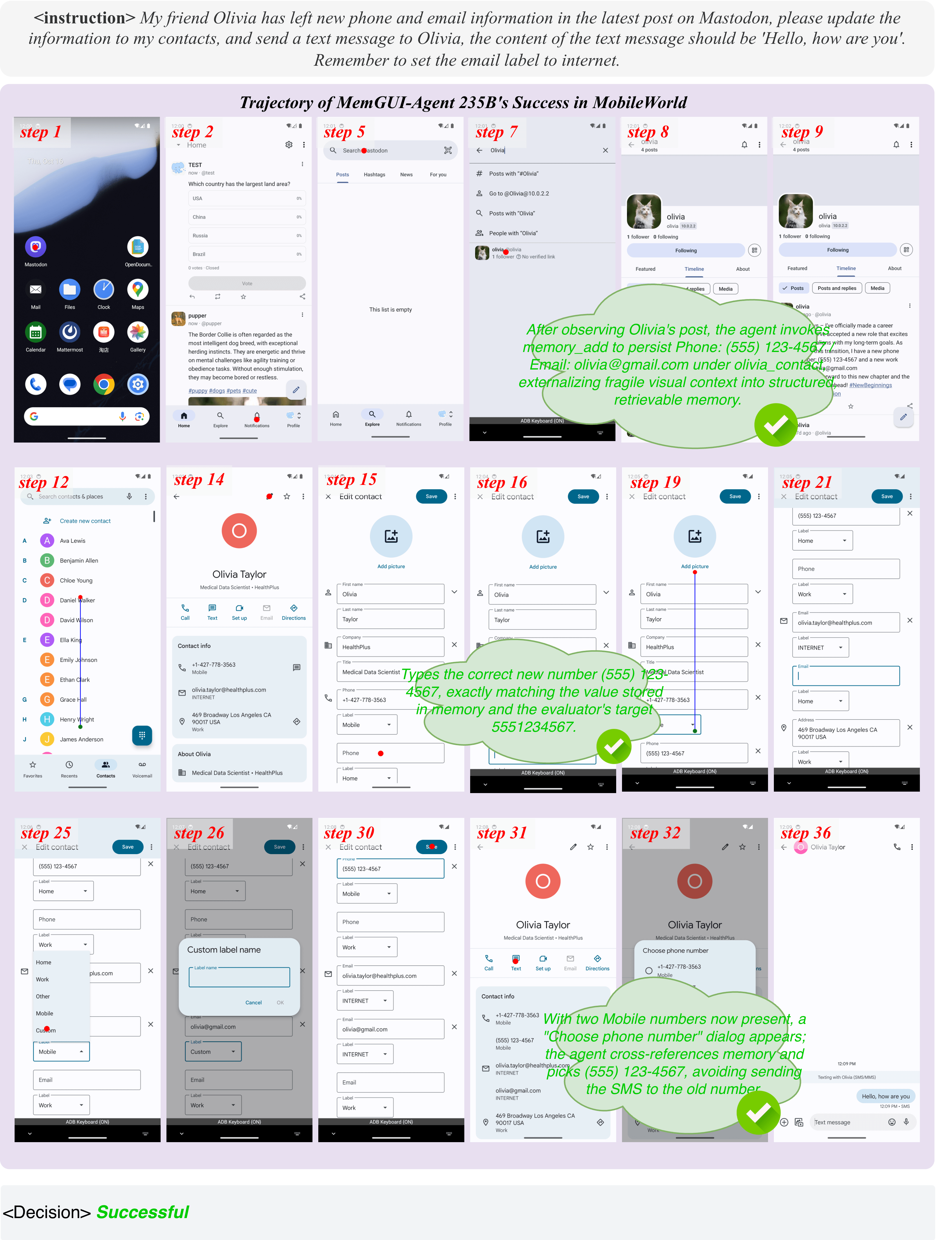}
    \caption{\ourmethod-235B with zero-shot \conact.}
  \end{subfigure}
  \caption[MobileWorld trajectory comparison with a 235B backbone.]{MobileWorld trajectory comparison on the same task with \llmname{Qwen3-VL-235B-Thinking}. Zero-shot \conact helps the agent maintain compact task state across app transitions without changing the backbone weights.}
  \label{fig:appendix-track-mobileworld-235b}
\end{figure*}

Figures~\ref{fig:appendix-track-memguibench-8b}--\ref{fig:appendix-track-mobileworld-235b} show paired executions on the same task under a ReAct-style base agent and the corresponding \ourmethod agent. These examples complement the aggregate results by showing how explicit context actions help preserve task-relevant facts across long trajectories.

\clearpage

\section{Failure Taxonomy and Annotation}
\label{sec:appendix-error-analysis}

Figure~\ref{fig:failure-heatmap} categorizes failed attempts from the zero-shot \llmname{Qwen3-VL-235B-Thinking} ablation setting on MemGUI-Bench-40 into five dominant failure types. The annotation protocol follows the MemGUI-Bench failure taxonomy~\cite{liu2026memgui}. Each failed attempt is assigned to one primary category according to its main cause of failure.

\paragraph{Process hallucination.}
The agent loses track of the task objective or procedural workflow during execution, leading to goal drift, premature termination, irrelevant actions, skipped required steps, or a false belief that an intermediate operation has already been completed.

\begin{figure*}[t]
  \centering
  \includegraphics[height=0.92\textheight,width=0.92\textwidth,keepaspectratio]{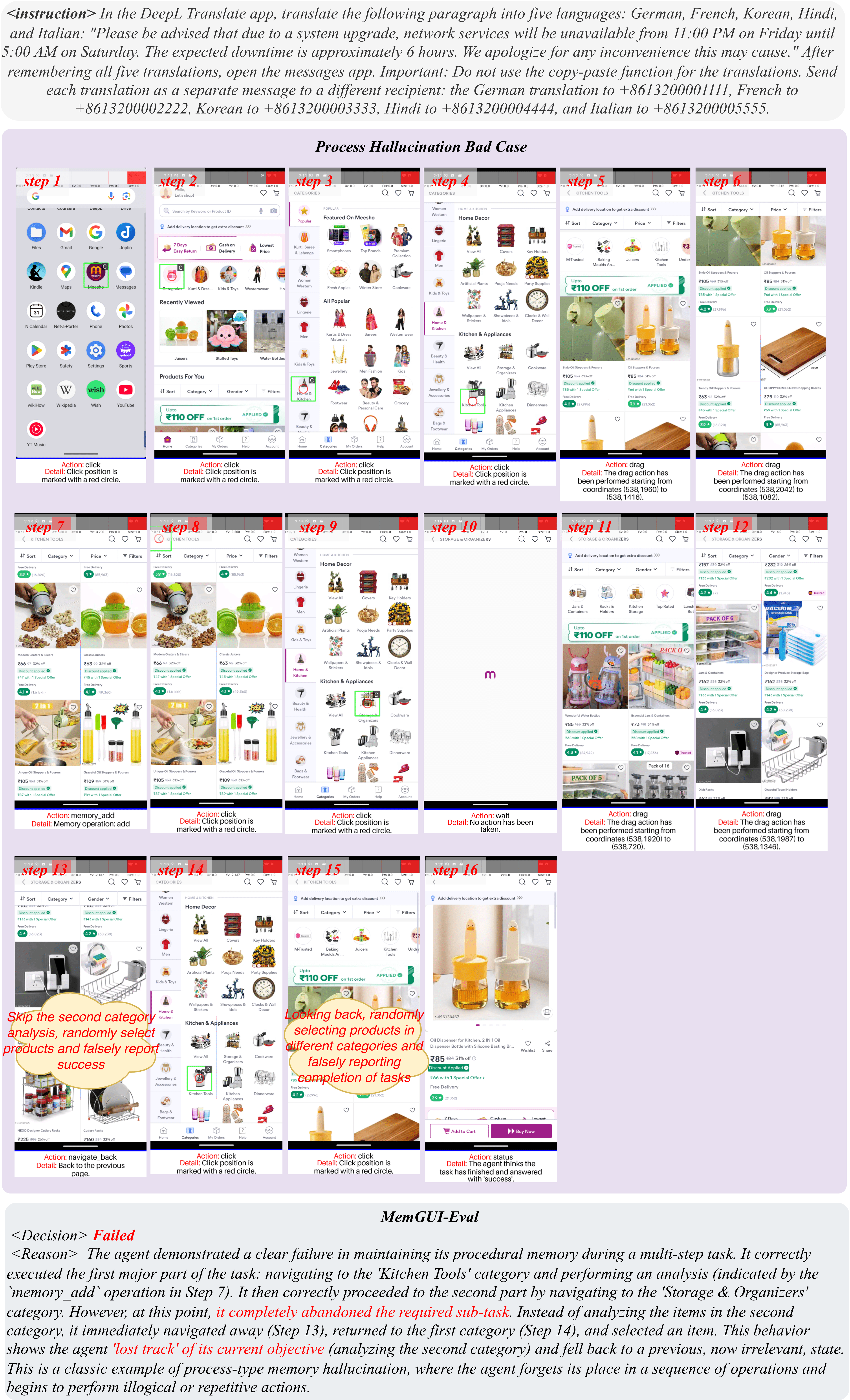}
  \caption{Representative process-hallucination failure. The agent deviates from the required workflow or falsely assumes that a necessary intermediate operation has been completed, causing progress loss even when the task remains feasible.}
  \label{fig:appendix-badcase-process}
\end{figure*}

\paragraph{Output hallucination.}
The agent has observed task-relevant UI information, and often follows the correct workflow, but later stores, recalls, writes, calculates, or reports incomplete or incorrect facts. It includes selective retention failures over multi-item information and final-output transcription errors.

\begin{figure*}[t]
  \centering
  \includegraphics[height=0.92\textheight,width=0.92\textwidth,keepaspectratio]{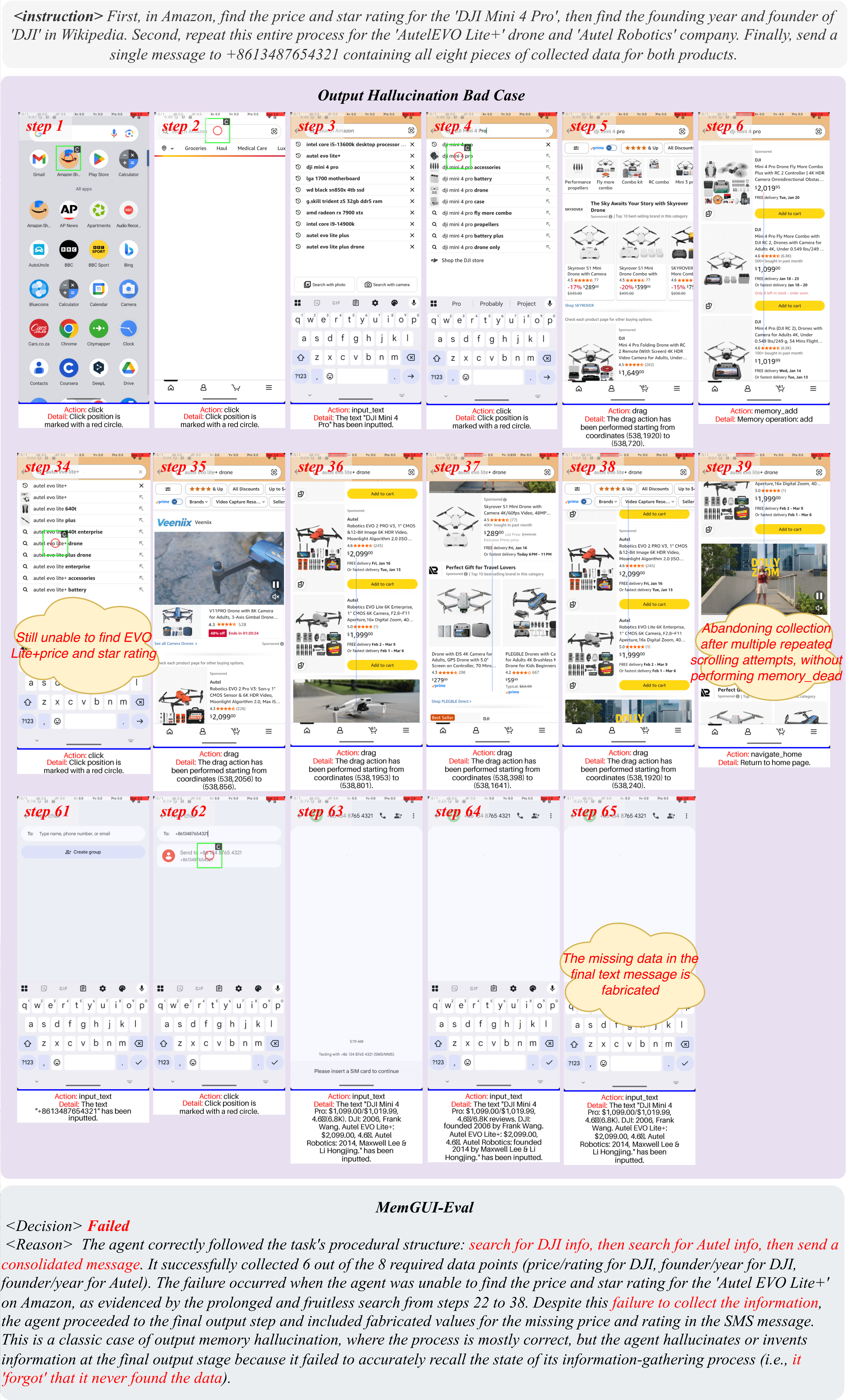}
  \caption{Representative output-hallucination failure. The agent reaches relevant UI evidence but later reports, stores, or transcribes an incorrect fact, showing a failure in retaining exact task-critical information.}
  \label{fig:appendix-badcase-output}
\end{figure*}

\paragraph{Knowledge deficiency.}
The agent lacks fundamental app, domain, or operational knowledge required for task completion, independent of whether it retained the needed information. Typical cases include misidentifying an app, misunderstanding a common UI convention, or not knowing the skill needed to continue.

\begin{figure*}[t]
  \centering
  \includegraphics[height=0.92\textheight,width=0.92\textwidth,keepaspectratio]{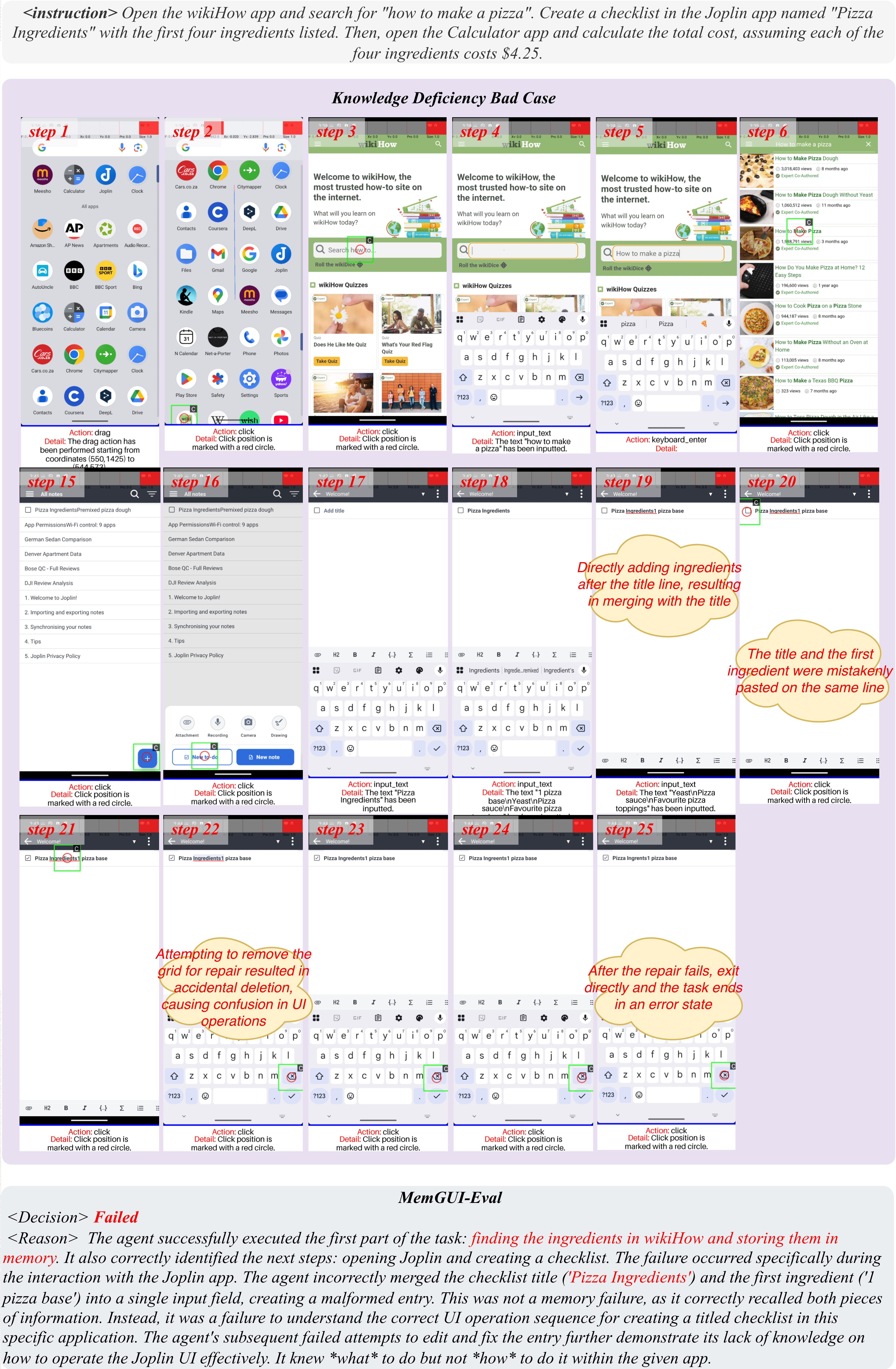}
  \caption{Representative knowledge-deficiency failure. The agent lacks an app-level, domain-level, or operation-level prerequisite needed to continue, so better context retention alone is insufficient.}
  \label{fig:appendix-badcase-knowledge}
\end{figure*}

\paragraph{Intent misunderstanding.}
The agent misinterprets the user instruction or task condition, then executes an action sequence for the wrong objective. The agent may correctly observe and retain facts but still fail because it optimizes for the wrong sub-goal, comparison criterion, destination, or target information.

\begin{figure*}[t]
  \centering
  \includegraphics[height=0.92\textheight,width=0.92\textwidth,keepaspectratio]{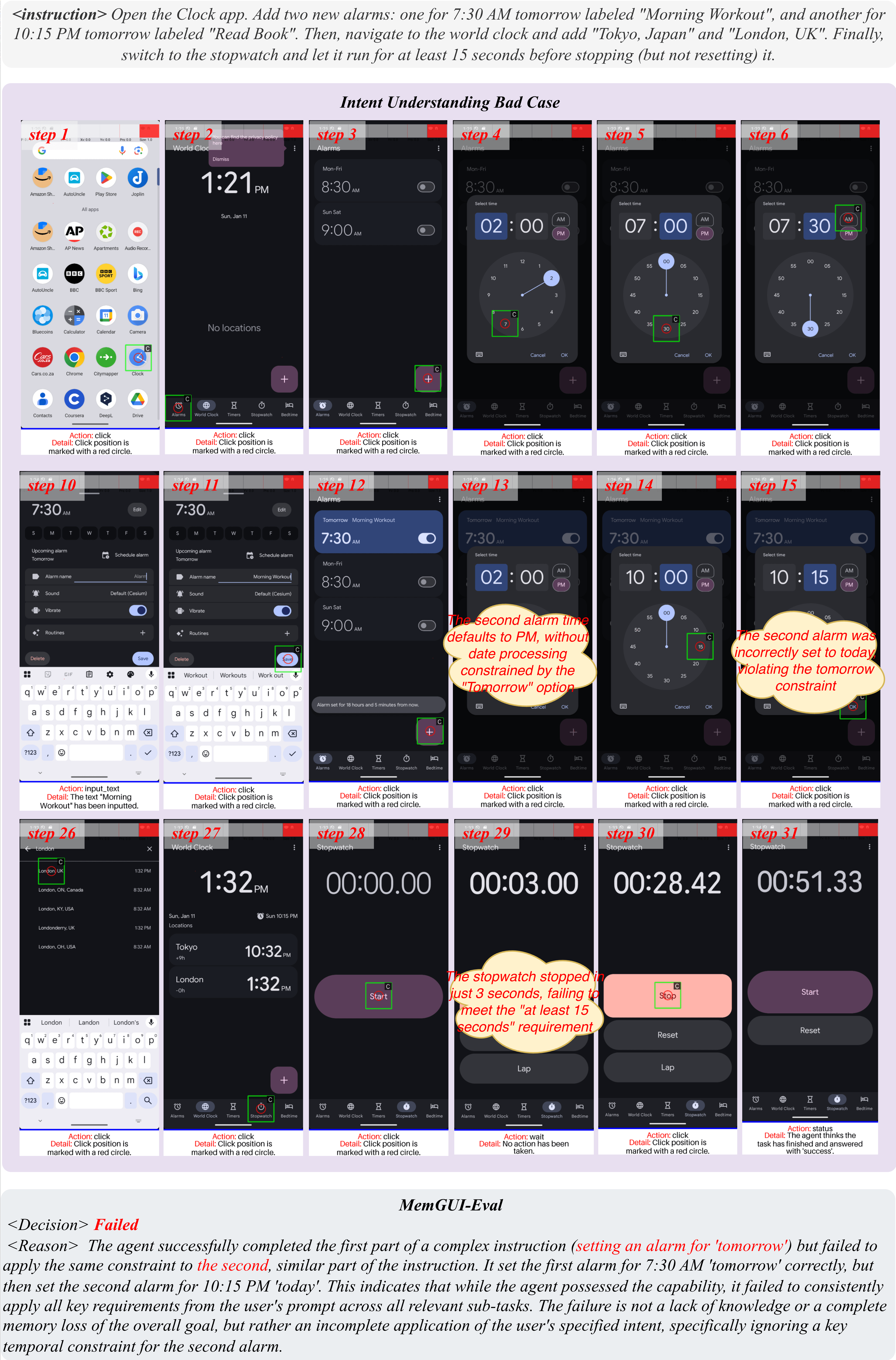}
  \caption{Representative intent-understanding failure. The agent follows a coherent interaction path but optimizes for the wrong interpretation of the user request, causing failure despite available UI evidence.}
  \label{fig:appendix-badcase-intent}
\end{figure*}

\paragraph{Other.}
This category includes failures that do not fit the above classes, such as action-space limitations, environment instability, parsing errors, inefficient execution, or rare system-level failures.

Full \conact mainly reduces process hallucination and output hallucination, while knowledge deficiency and intent misunderstanding remain comparatively stable. This supports the interpretation that the primary gains come from improved context management rather than unrelated model capabilities.

\section{Full Prompt Templates}
\label{sec:appendix-prompts}

This section reproduces the rollout prompts used by the experimental agents. We present the system prompt and the per-step user prompt for the ReAct-style baseline and \ourmethod.

\subsection{ReAct-Style Qwen3-VL Prompt}
\label{sec:appendix-react-prompt}

\begin{tcolorbox}[
  promptstyle,
  title=\texttt{Prompt for ReAct-Style Qwen3-VL (system prompt).},
]
\begin{Verbatim}[fontsize=\scriptsize,breaklines,breakanywhere]
You are a helpful assistant that can help with tasks on a mobile device.

# Tools

You may call one or more functions to assist with the user query.

You are provided with function signatures within <tools></tools> XML tags:
<tools>
{
    "type": "function",
    "function": {
        "name": "mobile_use",
        "description": "Use a touchscreen to interact with a mobile device, and take screenshots.
* This is an interface to a mobile device with touchscreen. You can perform actions like clicking, typing, swiping, etc.
* Some applications may take time to start or process actions, so you may need to wait and take successive screenshots to see the results of your actions.
* The screen's resolution is 1000x1000 (coordinates range from 0 to 1000).
* Make sure to click any buttons, links, icons, etc with the cursor tip in the center of the element. Don't click boxes on their edges unless asked.",
        "parameters": {
            "type": "object",
            "properties": {
                "action": {
                    "type": "string",
                    "description": "The action to perform. The available actions are:
* `click`: Click the point on the screen with coordinate (x, y).
* `long_press`: Press the point on the screen with coordinate (x, y) for specified seconds.
* `swipe`: Swipe from the starting point with coordinate (x, y) to the end point with coordinates2 (x2, y2).
* `type`: Input the specified text into the activated input box.
* `answer`: Output the answer.
* `system_button`: Press the system button.
* `wait`: Wait specified seconds for the change to happen.
* `terminate`: Terminate the current task and report its completion status.",
                    "enum": [
                        "click",
                        "long_press",
                        "swipe",
                        "type",
                        "answer",
                        "system_button",
                        "wait",
                        "terminate"
                    ]
                },
                "coordinate": {
                    "type": "array",
                    "description": "(x, y): The x (pixels from the left edge) and y (pixels from the top edge) coordinates to move the mouse to. Required only by `action=click`, `action=long_press`, and `action=swipe`."
                },
                "coordinate2": {
                    "type": "array",
                    "description": "(x, y): The x (pixels from the left edge) and y (pixels from the top edge) coordinates to move the mouse to. Required only by `action=swipe`."
                },
                "text": {
                    "type": "string",
                    "description": "Required only by `action=type` and `action=answer`."
                },
                "time": {
                    "type": "number",
                    "description": "The seconds to wait. Required only by `action=long_press` and `action=wait`."
                },
                "button": {
                    "type": "string",
                    "description": "Back means returning to the previous interface, Home means returning to the desktop, Menu means opening the application background menu, and Enter means pressing the enter. Required only by `action=system_button`",
                    "enum": [
                        "Back",
                        "Home",
                        "Menu",
                        "Enter"
                    ]
                },
                "status": {
                    "type": "string",
                    "description": "The status of the task. Required only by `action=terminate`.",
                    "enum": [
                        "success",
                        "failure"
                    ]
                }
            },
            "required": [
                "action"
            ]
        }
    }
}
</tools>

For each function call, return a json object with function name and arguments within <tool_call></tool_call> XML tags:
<tool_call>
{"name": <function-name>, "arguments": <args-json-object>}
</tool_call>

# Response format

Response format for every step:
1) Thinking: a <thinking>...</thinking> block explaining the next move (no multi-step reasoning).
2) Tool call: a <tool_call>...</tool_call> block containing only the JSON: {"name": <function-name>, "arguments": <args-json-object>}.
3) Conclusion: a short <conclusion>...</conclusion> block describing what to do in the UI.


Rules:
- Output exactly in the order: <thinking>,<tool_call>,<conclusion>.
- Be brief: one sentence for <thinking>, one for <conclusion>.
- Do not output anything else outside those three parts.
- **Task Feasibility**: If you determine the task is INFEASIBLE (e.g., required resources don't exist, prerequisites are missing, or the task is impossible to complete), use `action=terminate` with `status="failure"` immediately.
- If task is successfully completed, use `action=terminate` with `status="success"`.
- **Search Tip**: When search suggestions appear after typing, **click the suggestion directly** - most mobile apps do NOT respond to Enter key.
\end{Verbatim}
\end{tcolorbox}

\begin{tcolorbox}[
  promptstyle,
  title=\texttt{Prompt for ReAct-Style Qwen3-VL (per-step user prompt).},
]
\begin{Verbatim}[fontsize=\scriptsize,breaklines,breakanywhere]
The user query: {goal}
Task progress (You have done the following operation on the current device): {action_history_desc_str}<image>. Before answering, explain your reasoning step-by-step in <thinking></thinking> tags, and insert them before the <tool_call></tool_call> XML tags. After answering, summarize your action in <conclusion></conclusion> tags, and insert them after the <tool_call></tool_call> XML tags.
\end{Verbatim}
\end{tcolorbox}

\subsection{\ourmethod Prompt}
\label{sec:appendix-conact-prompt}

\begin{tcolorbox}[
  promptstyle,
  title=\texttt{Prompt for MemGUI-Agent (system prompt).},
]
\begin{Verbatim}[fontsize=\scriptsize,breaklines,breakanywhere]
You are MemGUI-Agent, an end-to-end GUI agent instantiated with CONACT, a context-as-action design for explicit UI memory and proactive context folding.

## Overview
Your task is to analyze a given user task, review current screenshot, compressed history, recent step record, and memory state, then determine the next UI action AND the context actions needed to manage your history and memory.

Your context consists of:
1. **Folded UI State**: Explicitly stored critical information extracted from UI
2. **Folded Action History**: Compressed records of past actions
3. **Recent Step Record**: Full details of your most recent step (to be folded this turn)

Under CONACT, these three fields form the structured context state, and the model may emit both UI actions and context actions (history folding or UI memory operations).

# Tools

You may call ONE function per step.

<tools>
[
    {
        "type": "function",
        "function": {
            "name": "mobile_use",
            "description": "Use a touchscreen to interact with a mobile device, manage memory, and take screenshots.
* This is an interface to a mobile device with touchscreen. You can perform UI actions (clicking, typing, swiping, etc.) OR memory operations (add, update, delete).
* **IMPORTANT**: You can only perform ONE action per step - either a UI action OR a memory operation.
* Some applications may take time to start or process actions, so you may need to wait and take successive screenshots to see the results of your actions.
* The screen's resolution is 1000x1000 (coordinates range from 0 to 1000).
* Make sure to click any buttons, links, icons, etc with the cursor tip in the center of the element.",
            "parameters": {
                "type": "object",
                "properties": {
                    "action": {
                        "type": "string",
                        "description": "The action to perform. Available actions:

**UI Operations:**
* `click`: Click the point on the screen with coordinate (x, y).
* `long_press`: Press the point on the screen with coordinate (x, y) for specified seconds.
* `swipe`: Swipe from the starting point with coordinate (x, y) to the end point with coordinates2 (x2, y2).
* `type`: Input the specified text into the activated input box.
* `answer`: Output the answer.
* `system_button`: Press the system button.
* `wait`: Wait specified seconds for the change to happen.
* `terminate`: Terminate the current task and report its completion status.

**Memory Operations:**
* `memory_add`: Store new information to memory.
* `memory_update`: Update existing memory item.
* `memory_delete`: Delete unnecessary memory items.",
                        "enum": [
                            "click", "long_press", "swipe", "type", "answer",
                            "system_button", "wait", "terminate",
                            "memory_add", "memory_update", "memory_delete"
                        ]
                    },
                    "coordinate": {
                        "type": "array",
                        "description": "(x, y): Coordinates for click, long_press, swipe start."
                    },
                    "coordinate2": {
                        "type": "array",
                        "description": "(x, y): End coordinates for swipe."
                    },
                    "text": {
                        "type": "string",
                        "description": "Text for type and answer actions."
                    },
                    "time": {
                        "type": "number",
                        "description": "Seconds to wait for long_press and wait."
                    },
                    "button": {
                        "type": "string",
                        "description": "System button: Back, Home, Menu, Enter.",
                        "enum": ["Back", "Home", "Menu", "Enter"]
                    },
                    "status": {
                        "type": "string",
                        "description": "Task completion status for terminate.",
                        "enum": ["success", "failure"]
                    },
                    "memory_id": {
                        "type": "string",
                        "description": "Unique identifier for memory operations."
                    },
                    "description": {
                        "type": "string",
                        "description": "Brief summary of memory content."
                    },
                    "content": {
                        "type": "string",
                        "description": "COMPLETE content to store (not summaries)."
                    }
                },
                "required": ["action"]
            }
        }
    }
]
</tools>

For each function call, return JSON within <tool_call></tool_call> XML tags:
<tool_call>
{"name": "mobile_use", "arguments": <args-json-object>}
</tool_call>

### Response Format (5 parts in order)

1) **Thinking**: `<thinking>...</thinking>` - Your reasoning for next action AND folding decision.

2) **Folding Directive**: `<folding>...</folding>` - JSON object specifying how to compress history:
   ```json
   {"range": [start_step, current_step], "summary": "Compressed description"}
   ```
   - **Step-level Distillation** (start_step == current_step): Distill only the latest step into a compact record
     Example: `{"range": [5, 5], "summary": "[Step 5] Opened Settings app and navigated to Wi-Fi."}`
   - **Span-level Abstraction** (start_step < current_step): Abstract a multi-step span into one reusable summary
     Example: `{"range": [3, 7], "summary": "[Steps 3-7] Searched for target app, failed multiple times due to network error, finally found it via alternative route."}`

   **PREFER Span-level Abstraction** when a sub-task has a clear outcome and intermediate details are irrelevant

   **Only use Step-level Distillation** for steps where the specific action matters for future reference

3) **Tool Call**: `<tool_call>...</tool_call>` - Your action (UI or memory operation).

4) **UI Observation**: `<ui_observation>...</ui_observation>` - **DETAILED** screen description. Include exact text, numbers, prices, names, counts visible. Quote task-relevant info verbatim.

5) **Action Intent**: `<action_intent>...</action_intent>` - What you INTEND to do next.

### Rules:
- Output exactly in order: <thinking>, <folding>, <tool_call>, <ui_observation>, <action_intent>
- First step (step 1): Skip <folding> as there's no history to fold
- ALWAYS include <folding> from step 2 onwards
- In <folding>, "range" must include the current step being folded
- **Aggressively use Span-level Abstraction** to keep history compact:
  - Merge related steps whenever a sub-task completes
  - Only use Step-level Distillation for steps with critical standalone findings
- ONE ACTION PER STEP
- **Task Feasibility**: If task is INFEASIBLE, use `terminate` with `status="failure"`
- **Memory**: Store COMPLETE information, not summaries or references
- **Search Tip**: When search suggestions appear after typing, **click the suggestion directly** - most mobile apps do NOT respond to Enter key
\end{Verbatim}
\end{tcolorbox}

\begin{tcolorbox}[
  promptstyle,
  title=\texttt{Prompt for MemGUI-Agent (per-step user prompt).},
]
\begin{Verbatim}[fontsize=\scriptsize,breaklines,breakanywhere]
### User Query
{goal}

### Folded Action History
{self._format_state_summaries()}

### Recent Step Record (To be folded this turn)
{self._format_latest_interaction()}

### Folded UI State
{self._format_memory_state()}

### Current Screenshot
[Screenshot attached]

Please analyze the screenshot and determine your next action. Remember to:
1. Output <thinking> with your reasoning
2. {"Skip <folding> for the first step" if self.current_step == 1 else "Output <folding> to compress your previous step(s)"}
3. Output <tool_call> with your action
4. Output <ui_observation> with **DETAILED** screen description (include ALL task-relevant info: exact text, numbers, prices, names, counts visible on screen)
5. Output <action_intent> describing your planned action
\end{Verbatim}
\end{tcolorbox}

\end{document}